\documentclass[12pt]{article}

\usepackage{latexsym,epsfig,graphicx,epstopdf,amsmath,amssymb,amscd}
\usepackage{multirow,paralist,dsfont,url}
\usepackage[titletoc]{appendix}
\usepackage[american]{babel}
\usepackage{xcolor}
\usepackage{grffile}

\usepackage{natbib}  

\newcommand{\thetavec}{{\boldsymbol{\theta}}}

\newcommand{\wvec}{{\boldsymbol{w}}}

\newcommand{\yvec}{{\boldsymbol{y}}}

\newcommand{\betavec}{{\boldsymbol{\beta}}}
\newcommand{\gammavec}{{\boldsymbol{\gamma}}}
\newcommand{\etavec}{{\boldsymbol{\eta}}}

\newcommand{\degreeC}{\ensuremath{^\circ}\text{C}}

\newcommand{\pr}{{\rm Pr}}

\newcommand{\NOR}{{\rm N}}

\newcommand{\thetavechat}{\widehat{\thetavec}}

\newcommand{\nor}{\textrm{nor}}

\newcommand{\Xvec}{\boldsymbol{X}}

\newcommand{\xvec}{\boldsymbol{x}}

\newcommand{\muvec}{\boldsymbol{\mu}}

\newcommand{\alphavec}{\boldsymbol{\alpha}}

\newcommand{\N}{\textrm{N}}

\newcommand{\D}{\mathcal{D}}
\newcommand{\thres}{\D_{0}}

\newcommand{\temp}{\texttt{temp}}
\newcommand{\ti}{\text{TI}}

\newcommand{\sigmaeps}{\sigma_{\epsilon}}

\def\baselinestretch{1.25}

\begin{document}

\title{What Quality Engineers Need to Know about Degradation Models}

\author{
Jared M. Clark$^{1}$\footnote{Corresponding Author. Email: cjared96@vt.edu}, Jie Min$^{2}$, Mingyang Li$^{3}$, Richard L. Warr$^{4}$,\\
Stephanie P. DeHart$^{1}$, Caleb B. King$^{5}$, Lu Lu$^{2}$, and Yili Hong$^{1}$\\[1.5ex]
{\small $^{1}$Department of Statistics, Virginia Tech, Blacksburg, VA}\\[0.05ex]
{\small $^2$Department of Mathematics \& Statistics, University of South Florida, Tampa, FL}\\[0.05ex]
{\small $^3$Department of IMSE, University of South Florida, Tampa, FL}\\[0.05ex]
{\small $^{4}$Department of Statistics, Brigham Young University, Provo, UT}\\[0.05ex]
{\small $^{5}$JMP Division, SAS, Cary, NC}
}

\date{}

\maketitle
\begin{abstract}
Degradation models play a critical role in quality engineering by enabling the assessment and prediction of system reliability based on data. The objective of this paper is to provide an accessible introduction to degradation models. We explore commonly used degradation data types, including repeated measures degradation data and accelerated destructive degradation test data, and review modeling approaches such as general path models and stochastic process models. Key inference problems, including reliability estimation and prediction, are addressed. Applications across diverse fields, including material science, renewable energy, civil engineering, aerospace, and pharmaceuticals, illustrate the broad impact of degradation models in industry. We also discuss best practices for quality engineers, software implementations, and challenges in applying these models. This paper aims to provide quality engineers with a foundational understanding of degradation models, equipping them with the knowledge necessary to apply these techniques effectively in real-world scenarios.


\end{abstract}

\newpage

\section{Introduction}\label{sec:introudction}

The development and large-scale deployment of new products, technologies, and materials require demonstrating their reliability, which we define as the ability to ensure confident use over time. This process involves collecting reliability data, as well as modeling and analyzing it. The two primary types of data used to assess reliability are time-to-event data and degradation data (e.g., \citealt{meeker2021statistical}). In this paper, we provide a comprehensive and accessible introduction to degradation data and its modeling and analysis. Degradation models are essential tools in reliability engineering and quality control, offering insights into the progressive deterioration of products or systems. In this section, we introduce key concepts that quality engineers need to understand about degradation models.

As noted by \citet{Condra1993}, reliability can be viewed as ``quality over time.'' Throughout a product's lifetime, its performance may gradually deteriorate, eventually reaching a point where it no longer functions as intended. This phenomenon is known as performance degradation, and the resulting failure is referred to as a soft failure. For example, the tread depth of a tire diminishes over time; once it falls below a certain threshold, the tire is no longer safe to use and should be replaced. During this process, degradation measurements are collected for each unit. Degradation models are statistical tools developed to model, analyze, and predict such deterioration over time using the collected data. Since degradation can often be observed before a product fails, these models enable reliability assessment even when data contain few or no failures.

Two main concepts in degradation modeling are the degradation path and the failure threshold. The degradation path represents how a performance characteristic changes over time, while the failure threshold defines the level of degradation at which the system is considered to have failed. Degradation measurements provide observed data points along this path. By modeling degradation progression and estimating when cumulative degradation is likely to cross the failure threshold, degradation analysis methods offer earlier and often more informative insights into product life than typical failure-time analysis, which are methods that rely solely on recorded failure-times and disregard any degradation information. The main idea of degradation modeling will be discussed in detail in Section~\ref{sec:degradation.data} and the sections that follow.

\begin{figure}
	\centering
	\includegraphics[width=.45\textwidth]{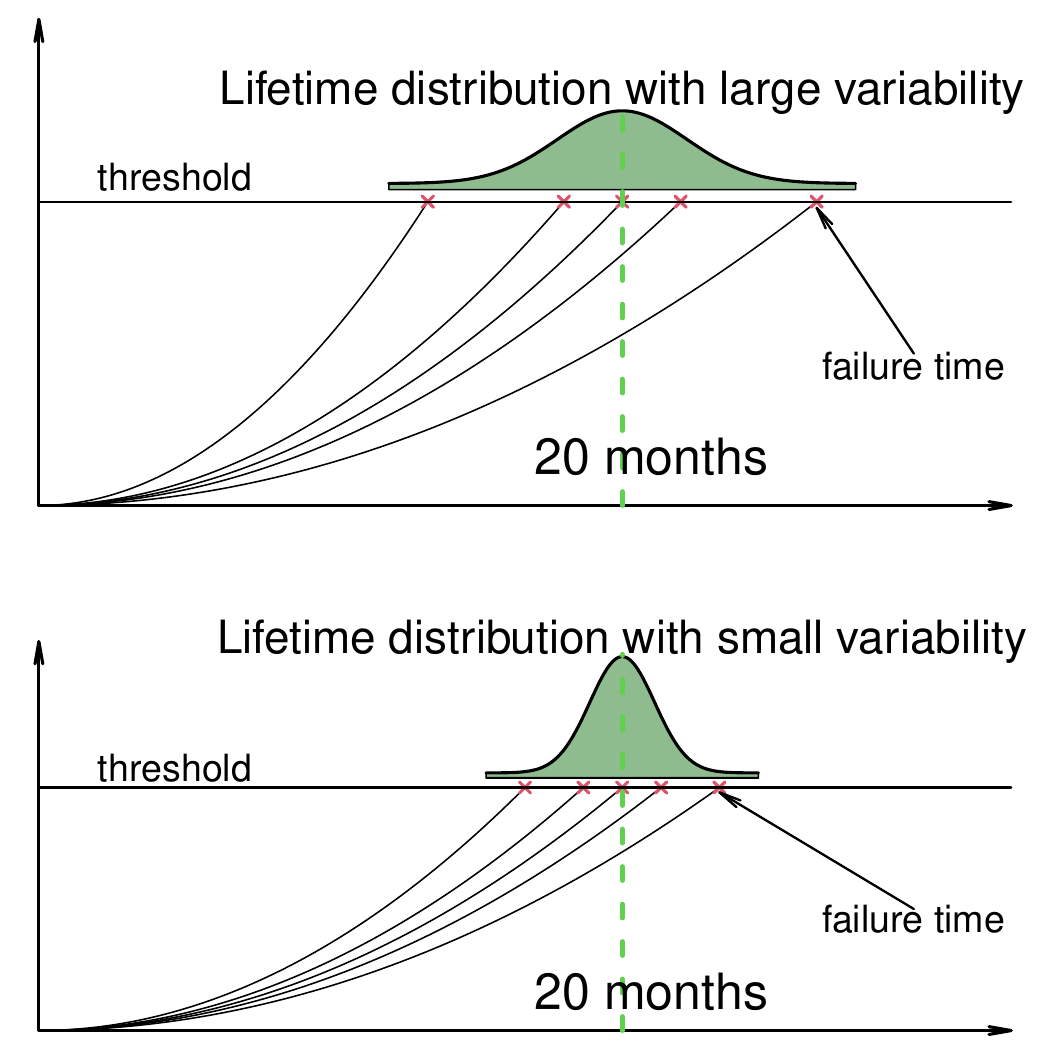}
	\caption{Illustration of the degradation paths and the failure threshold, which together induce a failure-time distribution.}\label{fig:dist}
\end{figure}

The reliability of products, systems, and materials is often of critical interest across various applications. Degradation models have been widely adopted in numerous industries and sectors which include aerospace, asset management, automotive, civil engineering, coatings, electronics, manufacturing, materials science, pharmaceuticals, and solar energy (e.g., \citealt{zhai2024modeling}, \citealt{bi2024degradation}, \citealt{Zhai03072023},  \citealt{nguyen2023bayesian}, \citealt{lu2021general}, \citealt{7803533}). In these fields, degradation modeling supports decision-making related to product design, maintenance planning, warranty analysis, and life cycle management.

Degradation analysis often involves collaboration among various professionals, including quality practitioners, reliability analysts, and other stakeholders such as product designers, operations managers, statisticians, and scientists who understand the degradation mechanism. These individuals work together to ensure effective reliability assessment and decision-making. For example, quality engineers may be responsible for establishing proper data collection and monitoring protocols, while reliability analysts may develop and apply models to evaluate and predict product life and failure risk. Product designers and operations managers may then use the resulting insights to inform design improvements, maintenance planning, and risk management strategies.

Degradation models offer significant advantages in quality engineering by enabling early detection of performance decline, which supports predictive maintenance and extends product life. This is especially valuable when failures are rare or costly to observe, making degradation modeling a practical alternative to reliability methods that rely solely on failure data. With modern technologies increasingly equipped with sensors, organizations can cost-effectively monitor the condition of systems in the field by collecting degradation data in real time. By leveraging gradual performance measurements, degradation models provide more accurate and precise life predictions and facilitate timely interventions (\citealt{lu1996comparison}). We focus on how degradation data can be used to assess reliability. Major questions when working with degradation data include how to collect the degradation measurements, which models to use for analysis, how to make inferences about the underlying degradation mechanisms and the failure-time distribution, and how to predict future degradation behavior. These aspects will be illustrated through multiple case studies in this paper.

A brief history of degradation data analysis can be summarized as follows. The concept of degradation as a measure of reliability originated in materials science and engineering. Early degradation studies primarily relied on failure time analysis, where degradation data were extrapolated to estimate pseudo failure times rather than directly modeling continuous degradation measurements. \citet{LuMeeker1993} were among the first to utilize repeated-measures degradation data for reliability assessment, proposing statistical models for analyzing degradation paths and demonstrating Monte Carlo methods for inference. Since then, degradation analysis has been widely adopted across diverse industries due to its numerous advantages. \citet{Meekeretal2014DegradationBookChapter} provided a comprehensive introduction to degradation data and related modeling methodologies. In parallel, various stochastic models have been developed, with a detailed review presented in \citet{YeXie2015DegradationReview}. Over time, degradation models have evolved to accommodate nonlinear trajectories, random parameters, and correlated measurements. Approaches such as general path models and stochastic process models have become widely used. Recent developments further integrate Bayesian methods, functional data analysis, and machine learning techniques to address complex data structures, including image data and sensory data.

The objective of this paper is to provide an accessible introduction to the fundamentals of degradation models, targeting quality practitioners, reliability analysts, and data scientists working in the quality industry. The rest of the paper is organized as follows. Section~\ref{sec:degradation.data} gives an overview of degradation analysis and provides a detailed introduction to various types of degradation data. Section~\ref{sec:degradation.model} presents the main statistical models used for degradation data analysis. Section~\ref{sec:inference.procedures} discusses inference procedures, including reliability estimation and prediction.
Section~\ref{sec:applications} presents case studies illustrating the use of degradation models in various application areas. Section~\ref{sec:borader.application.areas} highlights additional areas where degradation models have broader applications.
Finally, Section~\ref{sec:concluding.remarks} provides concluding remarks on practical considerations for applying degradation models, discusses software implementation and emerging trends in degradation modeling, and offers resources for further reading.

\section{Degradation Data}\label{sec:degradation.data}

 For a given system, we typically denote degradation at time $t$ as $\D(t)$, and we refer to the function, $\D(\cdot)$, as the degradation path. The degradation path is usually assumed to be a non-decreasing function of time, although in many scenarios degradation can be a non-increasing process, which can be addressed by considering $-\D(t)$ to ensure the path is non-decreasing. The non-decreasing assumption models the typical physical scenario that arises when damage on a system accumulates over time. Note that the phrase ``time'' is used in a broad sense here. If tire tread wear is measured to understand the reliability of tires, mileage could be the argument for the degradation path.

One typical goal in degradation analysis is to characterize the failure-time distribution. In order to do this with degradation data, we need to first define failure. When working with degradation data, a system experiences a ``soft'' failure once the degradation reaches a prespecified threshold. Let $\thres$ be our degradation threshold. Then the failure time $T$ is defined as,
\[ T = \min\{t:\D(t) \geq \thres\}.\]

Figure~\ref{fig:dist} illustrates the concepts of degradation path, failure threshold, and induced failure-time distribution. In this figure, the horizontal lines represent degradation thresholds. The curved lines are random realizations of degradation paths. Once a path crosses the threshold, a failure time is recorded which is a draw from the failure-time distribution. The distribution of failure-times relies heavily on the choice of threshold; imposing different thresholds on the same degradation paths could produce distributions with wildly different properties. Furthermore, the shape of the degradation paths will influence the distribution of failure times.

Degradation paths are considered random realizations of an underlying stochastic process. Given a failure threshold $\thres$, the degradation process induces a distribution of failure times. Observations of the degradation path $\D(t)$ are denoted by $y(t)$. A failure occurs when the degradation path $\D(t)$ crosses the threshold $\thres$ \citep{meeker2021statistical}.  The goal of modeling is to effectively capture the behavior of degradation data.

A number of datasets are introduced in the following sections. These datasets will be referenced throughout the paper. Table \ref{tab:data} provides a summary of all datasets discussed in this paper.

\begin{table}
\normalsize
    \caption{List of datasets used in this paper.}
  \begin{center}
    \label{tab:data}
    \begin{tabular}{c|c|c|c}\hline\hline
      Dataset Name &
      Data Type &
      Response &
      Introduction \\
      \hline
Laser Degradation & RMDT & Operating Current Increase & Section \ref{sec:RMDT.data} \\
Device B & RMDT & Power Drop & Section \ref{sec:RMDT.data}\\
Adhesive Bond B & ADDT & Bond Strength & Section \ref{sec:ADDT.data} \\
Outdoor Weathering & RMDT & Damage & Section \ref{sec:RMDT.data} \\
Coating & Dynamic covariates & Degradation & Section \ref{sec:coating} \\
Metal Fatigue & RMDT & Crack Length & Section \ref{sec:metal.fatigue} \\
Florida Roads & RMDT & Road Roughness & Section \ref{sec:civial.eng} \\\hline \hline
   \end{tabular}
  \end{center}
\end{table}

\subsection{Repeated-Measures Degradation Testing Data}\label{sec:RMDT.data}

Repeated measures degradation testing (RMDT) data arise when a time sequence of degradation measurements are recorded on the same unit. These measurements are non-destructive and do not affect the functionality of the unit, allowing repeated observations over real time. To introduce notation for RMDT data, suppose we have $n$ units, indexed by $i = 1, 2, \dots, n$. For each unit $i$, let $j = 1, 2, \dots, m_i$ index the replicate measurements, where $m_i$ is the number of measurements taken on unit $i$. The degradation measurement for unit $i$ at the $j$th recording time, $t_{ij}$, is denoted by $y_{ij}$. If accelerating variables or time-invariant independent variables are present, let $\xvec_i$ denote the vector of independent variable information associated with unit $i$.

\begin{figure}
\centering
\begin{tabular}{cc}
\includegraphics[width=0.465\textwidth]{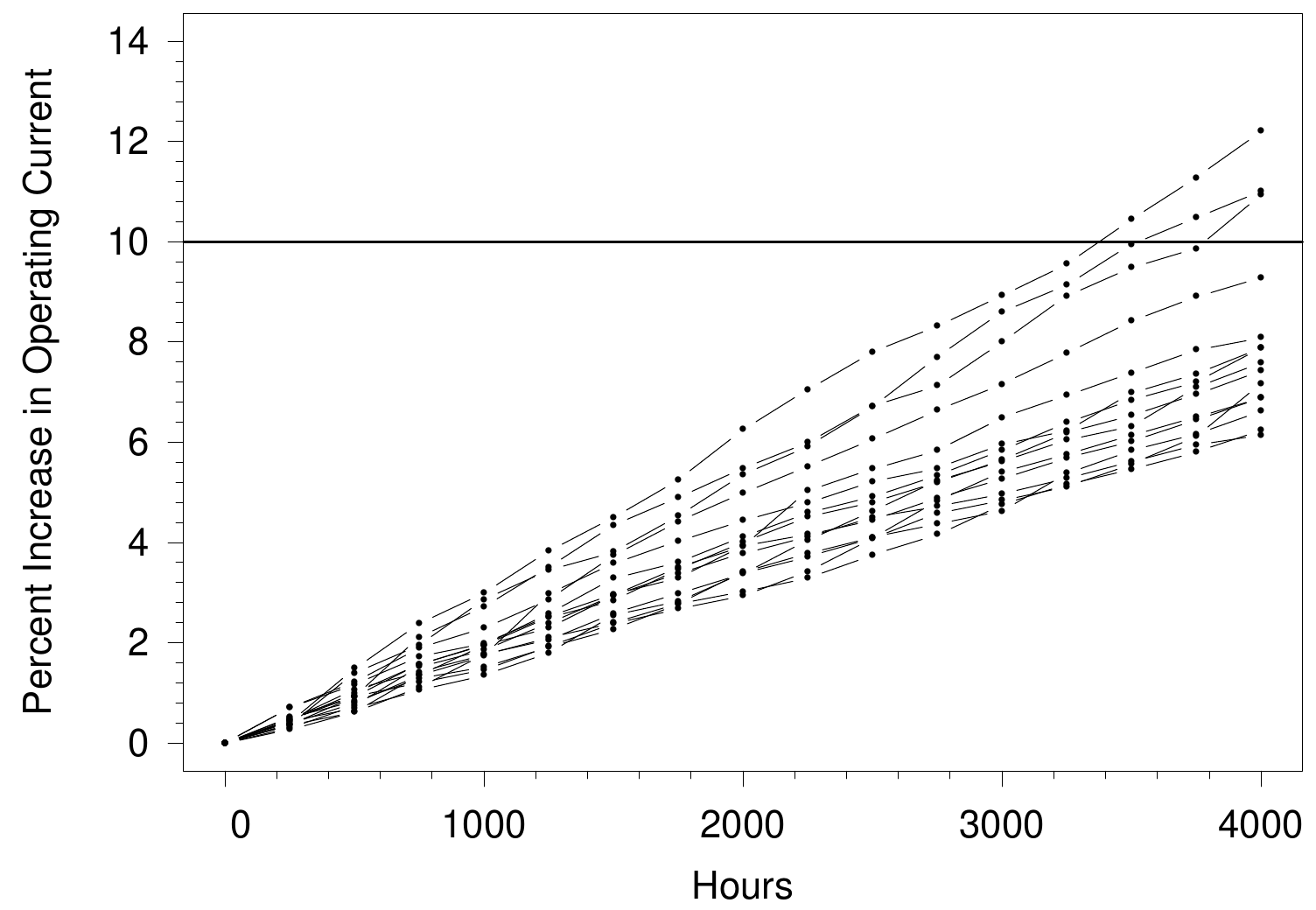}&
\includegraphics[width=0.48\textwidth]{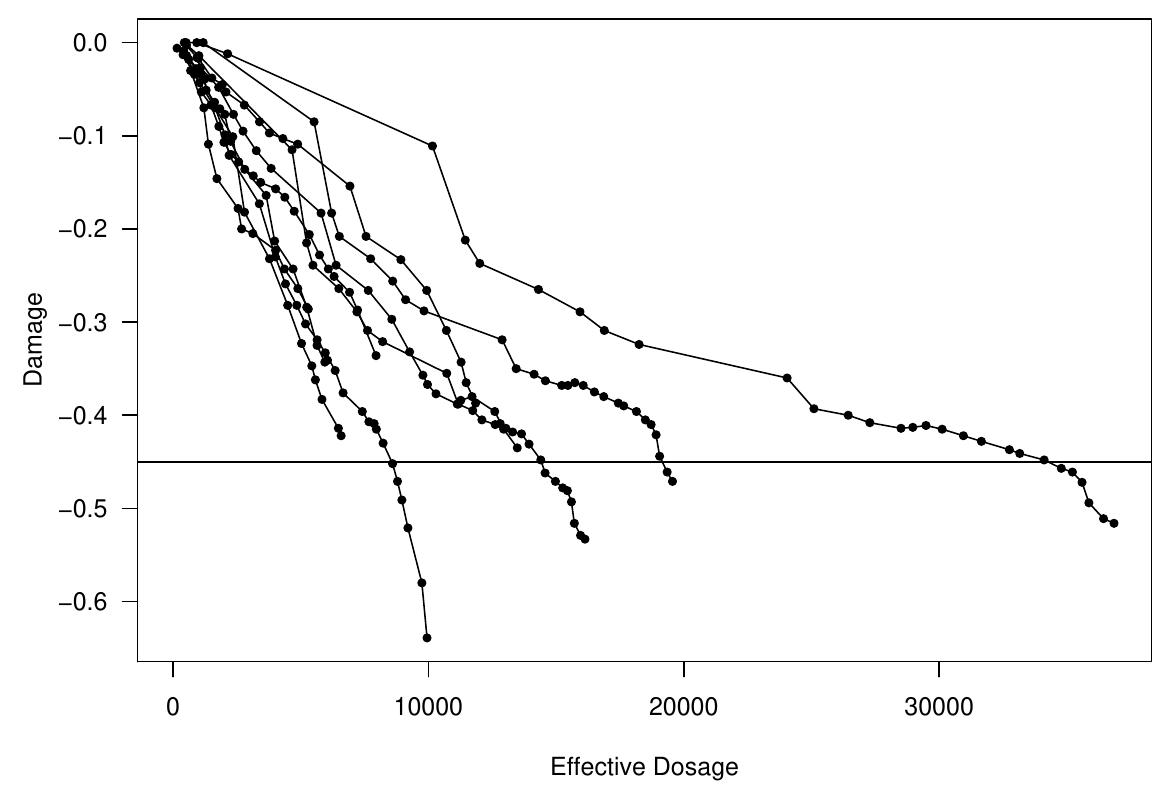}\\
(a) Linear Degradation & (b) Nonlinear Degradation
\end{tabular}
\caption{Illustration of degradation data: (a) Laser degradation data, an example of linear degradation; (b) Outdoor weathering data, an example of nonlinear degradation.
 \emph{Figure reproduced from Meeker et al.~(2011) with permission from Wiley.}}
\label{figure:laser.weathering.data}
\end{figure}

Figure \ref{figure:laser.weathering.data} illustrates RMDT data with two datasets from \cite{Meekeretal2014DegradationBookChapter}. The laser degradation data on the left provides an example of linear degradation, while the outdoor weathering data on the right exemplifies nonlinear degradation. The laser degradation data consists of repeated current measurements from laser units tested in a laboratory setting. These units, used in telecommunications, require increasing current over time to maintain constant light output, making operating current a degradation indicator. Figure~\ref{figure:laser.weathering.data}(a) shows the degradation paths for 15 tested units. Failure occurs when the units reach a 10\% increase in current which is represented by the horizontal line.  The outdoor weathering data provides photodegradation of an epoxy coating exposed to natural conditions at a certain location, with effective dosage as the relevant time scale. Figure~\ref{figure:laser.weathering.data}(b) shows degradation paths for a subset of specimens. A failure threshold of $-0.45$, following \citet{Meekeretal2014DegradationBookChapter}, is indicated by a horizontal line for illustration.

RMDT data may also involve accelerating variables and independent variables. Accelerated testing is a reliability evaluation technique used to speed up the degradation process, allowing observation of degradation paths within a practical time frame. Advances in engineering have led to highly durable products, making it impractical to conduct studies over their full life spans. To address this issue, accelerated tests expose units to elevated stress conditions that induce degradation or failure to occur more quickly, facilitating efficient data collection. Common accelerating variables include temperature, mechanical stress, voltage, ultraviolet (UV) radiation, humidity, and usage rate. A frequently used pedagogical example is temperature, which is applied at levels higher than typical use conditions to accelerate degradation.

\begin{figure}
\centering
\includegraphics[width=0.55\textwidth]{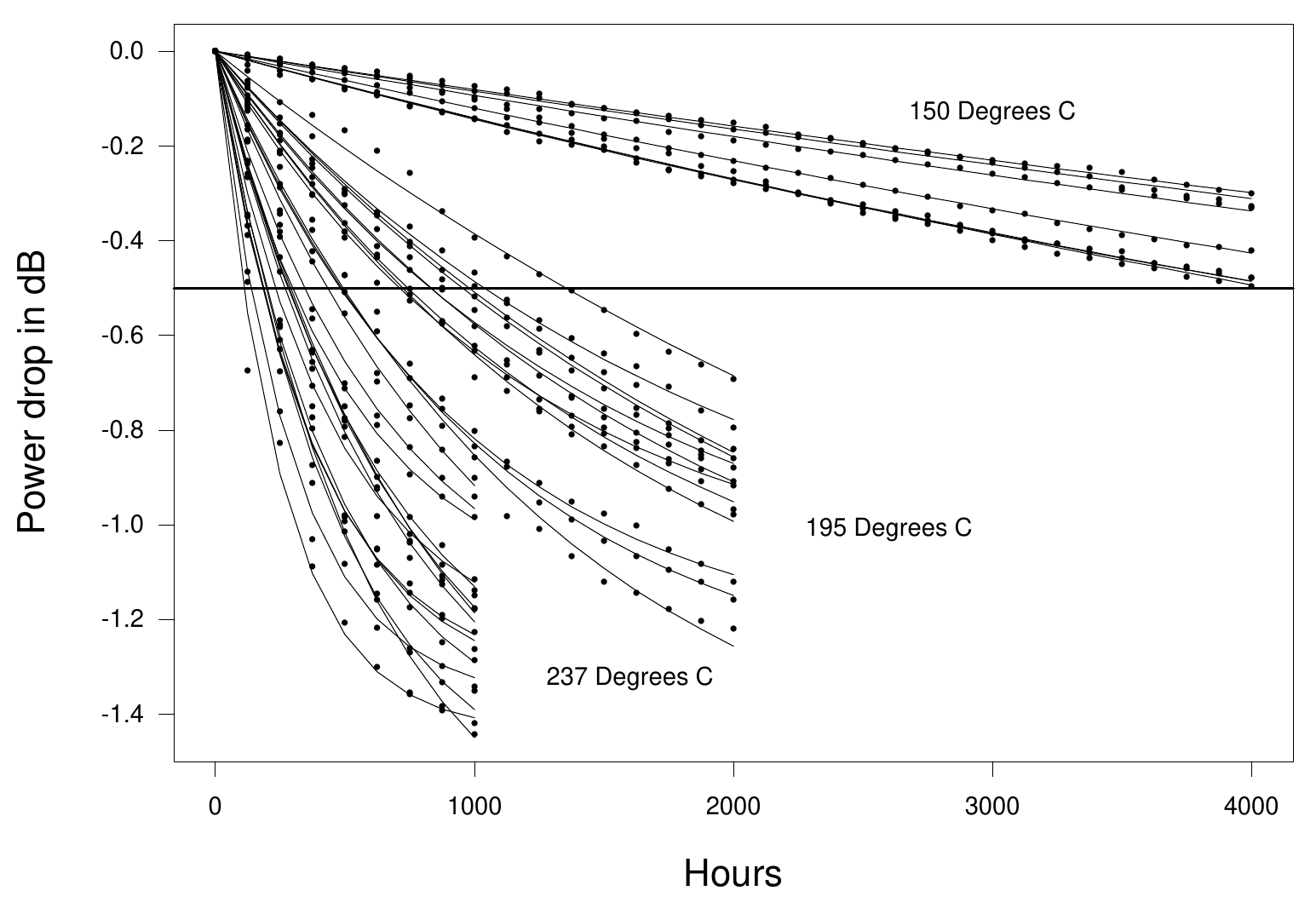}
\caption{Illustration of degradation data with an accelerating variable using the Device~B data and the fitted degradation paths. \emph{Figure reproduced from Meeker et al.~(2011) with permission from Wiley.}}
\label{figure:device.b.data}
\end{figure}

Figure~\ref{figure:device.b.data} illustrates the degradation paths of the Device~B data, which was introduced in \cite{MeekerEscobarLu1998} and \cite{meeker1998statistical} and was later featured in \citet{Meekeretal2014DegradationBookChapter},  with temperature as the accelerating variable. Device B is a radio-frequency power amplifier integrated circuit designed for normal operation at 80$\degreeC$. The accelerated RMDT plan involved testing devices at three levels of the accelerating variable temperature, 150$\degreeC$, 195$\degreeC$, and 237$\degreeC$, over a period of approximately six months. Failure for each unit was defined as the time when its output power declined by 0.5 decibels (dB) from its initial value.

Those examples shown in Figures~\ref{figure:laser.weathering.data} and~\ref{figure:device.b.data} illustrate that degradation paths can take various shapes, requiring different functional forms for modeling $\D(t)$. In addition, since multiple measurements are taken on each unit, the resulting observations are likely correlated, and this correlation should be accounted for in the modeling. Also, when accelerating variables or independent variables are present, they must be appropriately incorporated into the model.

With many of the examples in this paper, we see degradation data collected under indoor or laboratory conditions, where the accelerating variables are typically fixed over time. In contrast, degradation data from field tracking studies often include time-varying independent variables, which we refer to as dynamic independent variables. When an independent variable remains constant for a unit, it can be modeled as having a fixed effect on the degradation path. However, some independent variables, particularly those influencing the degradation path, vary over time and require more nuanced treatment.

Dynamic independent variables are common in many modern applications. For instance, the wear rate of a vehicle's finish may depend on UV exposure and temperature, both of which fluctuate over time for each individual vehicle in the study. In such cases, the independent variable exerts a cumulative effect depending on its time-varying trajectory. The increasing prevalence of dynamic independent variables is largely due to the widespread adoption of sensors, which can monitor system performance and usage conditions in real time.

\begin{figure}
\centering
\begin{tabular}{cc}
\includegraphics[width=.49\textwidth]{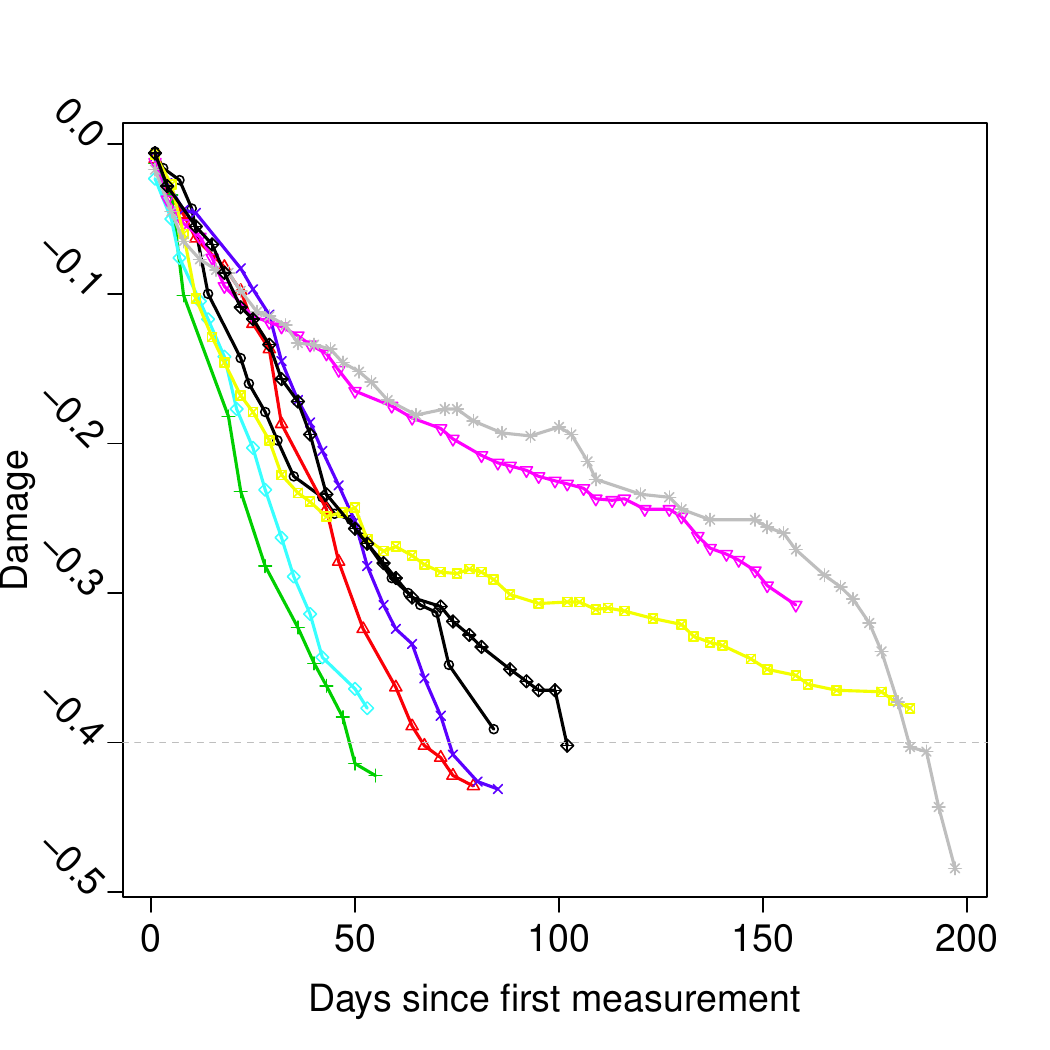}&
\includegraphics[width=.49\textwidth]{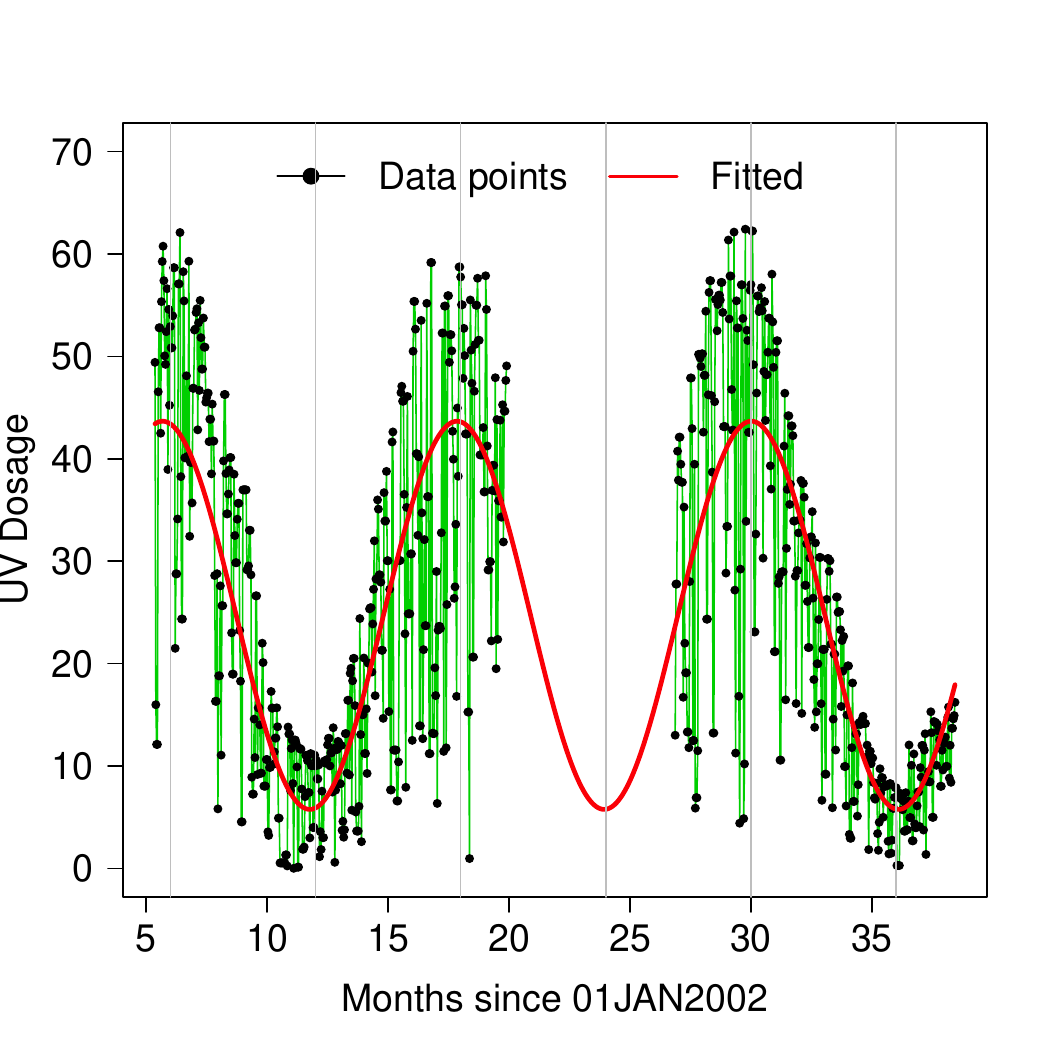}\\
(a) Degradation Path & (b) Daily UV Dosage
\end{tabular}
\caption{Plot of nine representative degradation paths (a) and corresponding dynamic independent variable information for daily UV dosage (b). Daily temperature and relative humidity data are not shown here. \emph{Figure reproduced from Hong et al.~(2015) with permission from Taylor and Francis.}}\label{fig:deg.cov.sample}
\end{figure}

We illustrate degradation data with dynamic independent variables using the outdoor weathering study from \citet{HongDuanetal2015}. To investigate organic coating degradation, thirty-six specimens  were studied in Gaithersburg, Maryland from 2002 to 2006. The specimens had different starting times over approximately five years. Sensors automatically recorded dynamic independent variables such as UV dosage, temperature, and relative humidity. Figure~\ref{fig:deg.cov.sample} shows nine representative degradation paths along with daily UV dosage. Temperature and relative humidity are omitted due to space limitation. We note, that while dosage is thought to be linked to the shape of the degradation path, this variable is changing over time.

\subsection{Accelerated Destructive Degradation Testing Data}\label{sec:ADDT.data}

Another type of degradation data encountered in practice is accelerated destructive degradation testing (ADDT) data. Unlike RMDT, the measurement process is destructive, and only one measurement can be taken on each unit. For example, measuring the tensile strength of a material requires pulling the specimen until it breaks. Due to the destructive nature of testing, the full degradation paths cannot be observed.

\begin{figure}
\begin{center}
\includegraphics[width=0.5\textwidth]{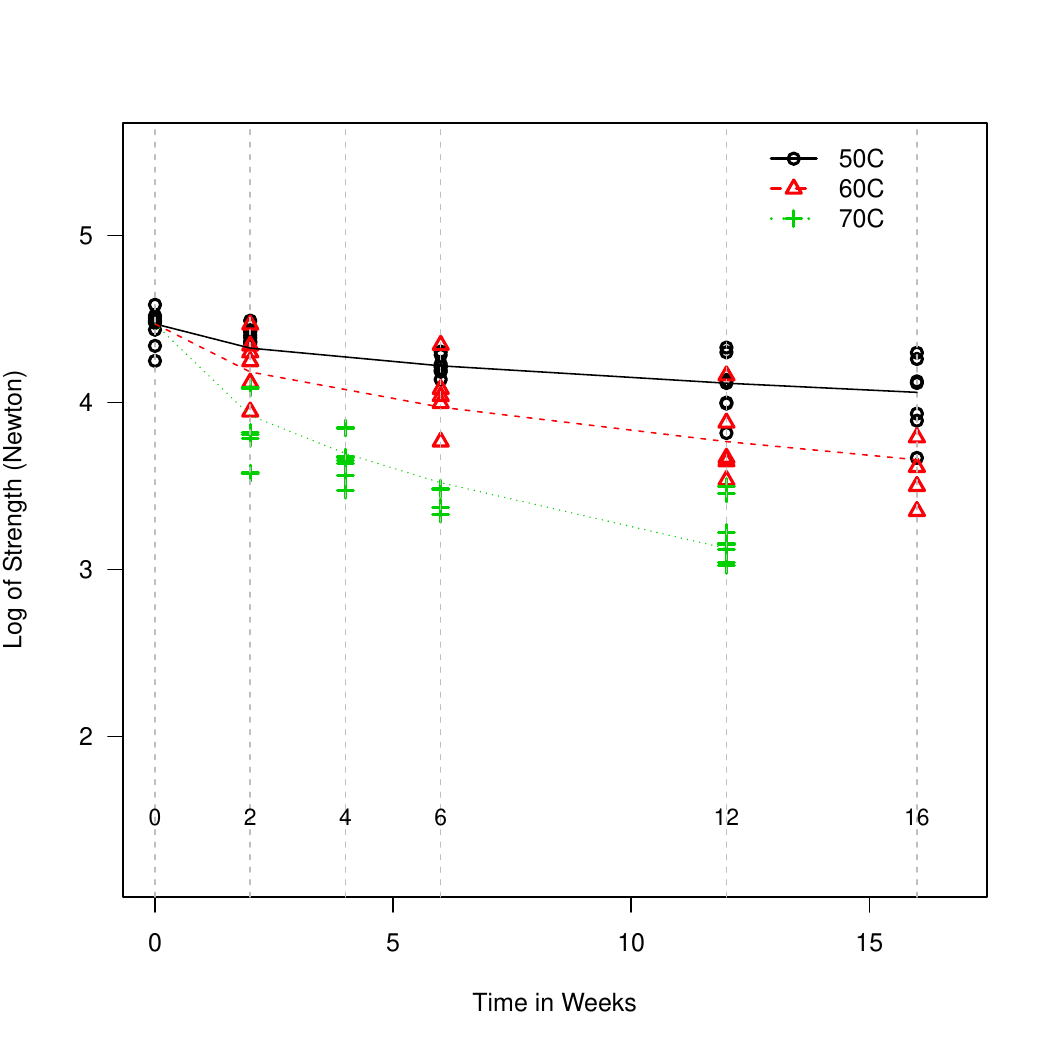}
\caption{Visualization of Adhesive Bond B data with scatter plot and fitted parametric degradation paths. \emph{Figure reproduced from Xie et al.~(2018) with permission from Taylor and Francis.}} \label{fig:AdhesiveBondB.data}
\end{center}
\end{figure}

\cite{Meekeretal2014DegradationBookChapter} provide an example of ADDT data with the Adhesive Bond B strength data, which was first introduced in \cite{EscobarMeekerKuglerKramer2003}. This data was also analyzed in \cite{meeker2021statistical} following a Bayesian approach. Figure~\ref{fig:AdhesiveBondB.data} shows a scatter plot of the Adhesive Bond B dataset where each observation corresponds to a distinct unit since strength could only be measured once per unit. Temperature served as the accelerating variable, with units exposed to three elevated temperature levels ($50\degreeC$, $60\degreeC$, and $70\degreeC$). At the start of the experiment, eight units were measured under normal temperature conditions to establish baseline strength. Subsequent measurements were taken at selected time points (weeks 2, 4, 6, 12, and 16) at each of the three temperature levels.

In modeling, the considerations for ADDT data differ from those for RMDT data, as we do not have multiple observations from the same unit and cannot distinguish variability across units from random measurement error. Additionally, samples measured at the same time point and under the same accelerating conditions tend to exhibit correlation. Therefore, special attention is required when modeling ADDT data.

To introduce notation for ADDT data, let $i$ index the test conditions, $j$ index the time points, and $k$ index the sample units. Let $t_j$ denote the time at which measurements are taken, and $x_i$ represents the test condition (e.g., temperature level). The degradation measurement $y_{ijk}$ denotes the response from unit $k$ measured at time $t_j$ under condition $x_i$.

Dynamic independent variables are an important consideration when a unit is observed over time. For this reason, they are most commonly encountered in RMDT data. However, in ADDT data, one may also have independent variable histories recorded up to the time of the destructive measurement. To introduce notation for dynamic independent variables, suppose there are $n$ units in the field. Let $n_i$ denote the number of measurement time points for unit $i$, and let $t_{ij}$, $j = 1, \dots, n_i$, denote the corresponding time points. The value of independent variable~$l$ for unit $i$ at time $s$ is denoted by $x_{il}(s)$. The independent variable history for unit $i$ up to time $t$ is represented by $\xvec_i(t) = \{x_i(s): 0 \leq s \leq t_{in_i}\}$, where $x_i(s) = [x_{i1}(s), \dots, x_{ip}(s)]'$.


\section{Degradation Models}\label{sec:degradation.model}

As with any type of data, new models for degradation data are continually in development. Since RMDT data are observed over time, many time series models can be applied to these data. Time series models treat measurements as correlated random variables which can be applicable when measurements are recorded on the same unit. Machine learning models are often useful as well. These models excel at prediction and many efforts are working towards increased interpretability and uncertainty quantification for these models. We acknowledge that not all degradation data can be easily categorized as RMDT or ADDT data. For example, degradation data may come in the form of functional data and require methods from functional data analysis. Recognizing that these approaches may be valuable in many cases, we limit our scope to general path models and stochastic process models for RMDT and ADDT data in this paper.

\subsection{General Path Models}

General path models (GPMs) take the following form:
\begin{align}\label{sec:GPM.general.form}
y(t) = \D(t) + \epsilon(t),
\end{align}
where $y(t)$ is the degradation measurement at time $t$, $\D(t)$ is the underlying degradation path, and $\epsilon(t)$ is the random error term. This modeling approach specifies a physically motivated parametric model for $\D(t)$, with some parameters treated as random to account for unit-to-unit variability. The model parameters are then estimated using the observed data.

Degradation paths are assumed to be monotonic functions of time and can be broadly classified as linear, concave, or convex. For certain path shapes, it is useful to express the model using differential equations. For example, the linear degradation model can be reformulated as:
\begin{align}\label{eqn:gpm.linear}
\frac{d\D(t)}{dt} = \theta.
\end{align}
Note that the linear model in \eqref{eqn:gpm.linear} involves two parameters: $\D(0)$ and $\theta$. Here, $\D(0)$ represents the initial degradation level at time $t=0$, and $\theta$ is the slope parameter that governs the rate of degradation accumulation. To account for unit-to-unit variability, these parameters can be indexed by unit.

A common convex degradation path model characterizing crack growth is the Paris-rule model (e.g., \citealt{Dowling1993}), which is defined by the following differential equation:
\[ \frac{d\D(t)}{dt} = \theta_1 \{f[\D(t)]\}^{\theta_2}. \]
In this model, $f(\cdot)$ is known as the stress intensity range function, and the solution to the differential equation depends on the specific form of $f$. This model incorporates two parameters, $\theta_1$ and $\theta_2$. One form of this model commonly used is $f[\D(t)] = x\sqrt{\pi\D(t)}$, where $x$ is proportional to the applied stress. Under this form, the solution to the differential equation is:
\begin{align}\label{eqn:paris.rule}
\D(t) & = \begin{cases}
      \{[\D(0)]^{1-\theta_2/2} + (1-\theta_2/2)\theta_1(x\sqrt{\pi})^{\theta_2}t\}^{2/(2-\theta_2)}, & \theta_2 \neq 2 \\
      \D(0)\exp[\theta_1(x\sqrt{\pi})^2t], & \theta_2 = 2
   \end{cases}.
\end{align}

Finally, we present an approach for modeling degradation paths using the log-logistic function. The model is defined by the following equation (e.g., \citealt{Fockeetal2017}):
\[ \frac{d\D(t)}{dt} = \frac{\alpha}{\kappa\nu}\left[\frac{\D(t)}{\alpha}\right]^{1-\kappa}\left[1-\frac{\D(t)}{\alpha}\right]^{1+\kappa},\]
which has one solution as,
\begin{align}\label{eqn:log.logistic.form}
\D(t) = \frac{\alpha}{1+(t/\nu)^{-1/\kappa}}.
\end{align}
The model in \eqref{eqn:log.logistic.form} provides the flexibility to capture a wide range of degradation path shapes, including concave and S-shaped curves and relies on the parameters $\alpha, \nu,$ and $\kappa$.

As mentioned earlier, practitioners use random effects to account for unit-to-unit variability. The general form of a mixed effects GPM in \eqref{sec:GPM.general.form} can thus be expressed as,
\begin{align}\label{eqn:degradation.path.ran.eff}
y_{ij}&=\D_{i}(t_{ij};\alphavec,\betavec_i)+\epsilon_{ij},
\end{align}
where $\alphavec$ is a vector of fixed parameters and $\betavec_i$ is a vector that represents unit-specific random effects. The random vector $\betavec_i$ is usually assumed to follow a multivariate normal distribution, $\NOR(\muvec_{\betavec}, \Sigma_{\betavec})$. The error term $\epsilon_{ij}$ is independently distributed as $\text{Normal}(0, \sigmaeps)$.

Assuming an increasing degradation path, the model in \eqref{eqn:degradation.path.ran.eff} leads to a failure time cumulative distribution function (CDF) given by given by
\begin{align}
F(t) = F(t; \thetavec) = \pr[\D(t; \alphavec, \betavec) \geq \thres].
\end{align}
Note that $\betavec$ is random and $F(t)$ depends only on the parameter vector $\thetavec$, which contains $\alphavec, \muvec_{\betavec}$, and $\Sigma_{\betavec}$.

\subsection{Stochastic Process Models}

We now turn to an alternative approach for modeling degradation using stochastic process (SP) models. The three most commonly used processes for modeling degradation paths are the Wiener process~\citep{Whitmore1995}, gamma process~\citep{LawlessCrowder2004}, and inverse Gaussian (IG) process~\citep{WangXu2010}. We begin with the Wiener process model, assuming the form
$$
y(t) = \mu(t) + \sigma W[\mu(t)],
$$
where $\mu(t)$ represents the mean degradation trend over time, $\sigma$ is a scale parameter, and $W(\cdot)$ denotes a standard Wiener process. Define the increment $\Delta_{21} = W[\mu(t_2)] - W[\mu(t_1)]$ for $t_2 > t_1$, which satisfies
$\Delta_{21} \sim \text{Normal}(0, \sqrt{\mu(t_2) - \mu(t_1)}),$
characterizing the Wiener process structure. We note the independent increment structure. This model leads to a transformed IG distribution for the failure time.  Specifically, let $T$ denote the first hitting time for a failure threshold $\thres$. We have $T = \mu^{-1}(T^{\ast})$, where
$$
T^{\ast} \sim \text{IG}(\thres, \thres^2/\sigma^2).
$$

Another important SP model is the gamma process, which also exhibits independent increments. As before, we use a general trend function $\mu(t)$. Under the gamma process, the increment $\D(t_2) - \D(t_1)$ follows a gamma distribution with shape parameter $\mu(t_2) - \mu(t_1)$ and an unknown scale parameter $\sigma$. Let $G(t; \alpha, \beta)$ denote the CDF of the gamma distribution with shape $\alpha$ and scale $\beta$. Then, the CDF of the failure time is given by
$$
F(t) = 1 - G[\thres; \mu(t), \sigma].
$$

The final SP model we consider is the inverse Gaussian process for modeling degradation data, which also exhibits the independent increment property. Let $\text{IG}(t; \mu, \sigma)$ denote the CDF of the inverse Gaussian distribution with mean $\mu$ and scale $\sigma$. Then, under the IG process model, the CDF of the failure time is given by
$$
F(t) = 1 - \text{IG}[\thres; \mu(t), \sigma \mu^2(t)].
$$

We recognize a number of considerations when choosing between the GPM and SP model approaches. The GPM is a flexible approach applicable to various types of degradation data, including RMDT and ADDT. It can readily incorporate physical knowledge into the mean trend function and can easily include independent variables, even dynamic ones. The GPM is intuitive because it models the degradation path directly and is implemented in several commercial software packages. For SP models, a closed-form expression for the induced CDF can be obtained under simple mean trend functions. However, SP models have limitations when dealing with nonlinear mean trend function, dynamic independent variables, and ADDT data. Especially, modeling becomes more challenging when the degradation trend is nonlinear.

In terms of inference, the confidence bands of the CDF may become narrow when the error variance is large under GPMs. Stochastic process models, by contrast, treat those deviations (the error terms in the GPM) as part of the distribution and therefore better capture uncertainty. However, if the mean trend is misspecified, the stochastic model results can be biased. Overall, based on our experience and the diverse examples considered in this paper, our recommendation is to use GPM in most cases. Further discussion of these two modeling approaches can be found in \cite{YeXie2015DegradationReview}.

The remainder of the paper will focus on GPMs as a practical approach to degradation modeling. For those looking for more details on SP models, \citet{YeXie2015DegradationReview} provide a comprehensive review of SP models. \citet{Peng2016} discusses the extension of SP models to incorporate random effects and measurement errors.

\section{Inference Procedures}\label{sec:inference.procedures}
\subsection{General Path Models for RMDT Data}

Maximum likelihood methods may be used to estimate the parameters governing the data-generating process. The likelihood function corresponding to model~\eqref{eqn:degradation.path.ran.eff} is given by
\begin{align}\label{eqn:lik.data}
L(\thetavec,\sigmaeps^2|\text{Data}) = \prod_{i=1}^n \int_{\betavec_i}
\left[\prod_{j=1}^{m_i} \frac{1}{\sigmaeps} \phi_{\nor}(z_{ij})\right]
f_{\betavec}(\betavec_i;\muvec_{\betavec},\Sigma_{\betavec}) d\betavec_i,
\end{align}
where $z_{ij} = [y_{ij} - \D(t_{ij}; \alphavec, \betavec_i)] / \sigmaeps$, $\phi_{\nor}$ denotes the standard normal probability density function (PDF), and $f_{\betavec}(\cdot)$ is the multivariate normal PDF.

The maximum likelihood estimates of $\thetavec$ and $\sigmaeps$ are obtained by maximizing the likelihood function in \eqref{eqn:lik.data}. The model can be fit using the \texttt{nlme} package in R~\citep{pinheiro2006nlme}. \citet{Meekeretal2014DegradationBookChapter} provides two algorithms which can be used to estimate $F(t)$ and compute the corresponding confidence intervals (CIs).

\begin{figure}
\centering
\begin{tabular}{cc}
\includegraphics[width=0.48\textwidth]{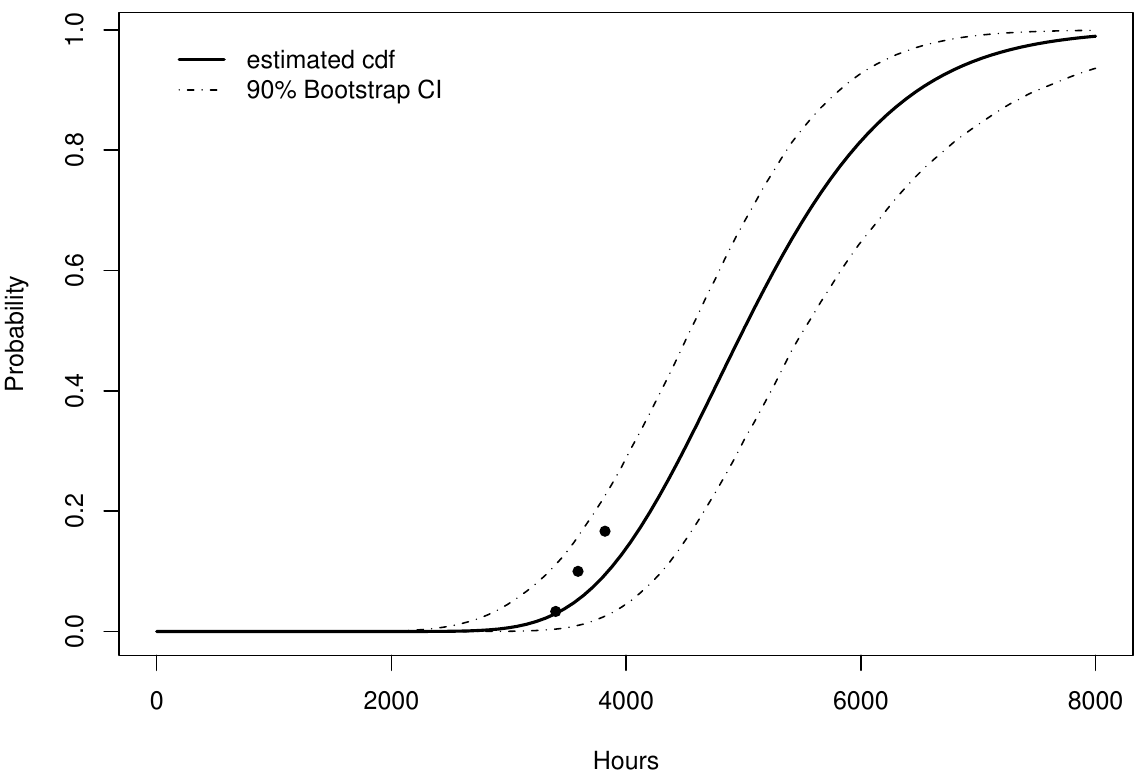}&
\includegraphics[width=0.48\textwidth]{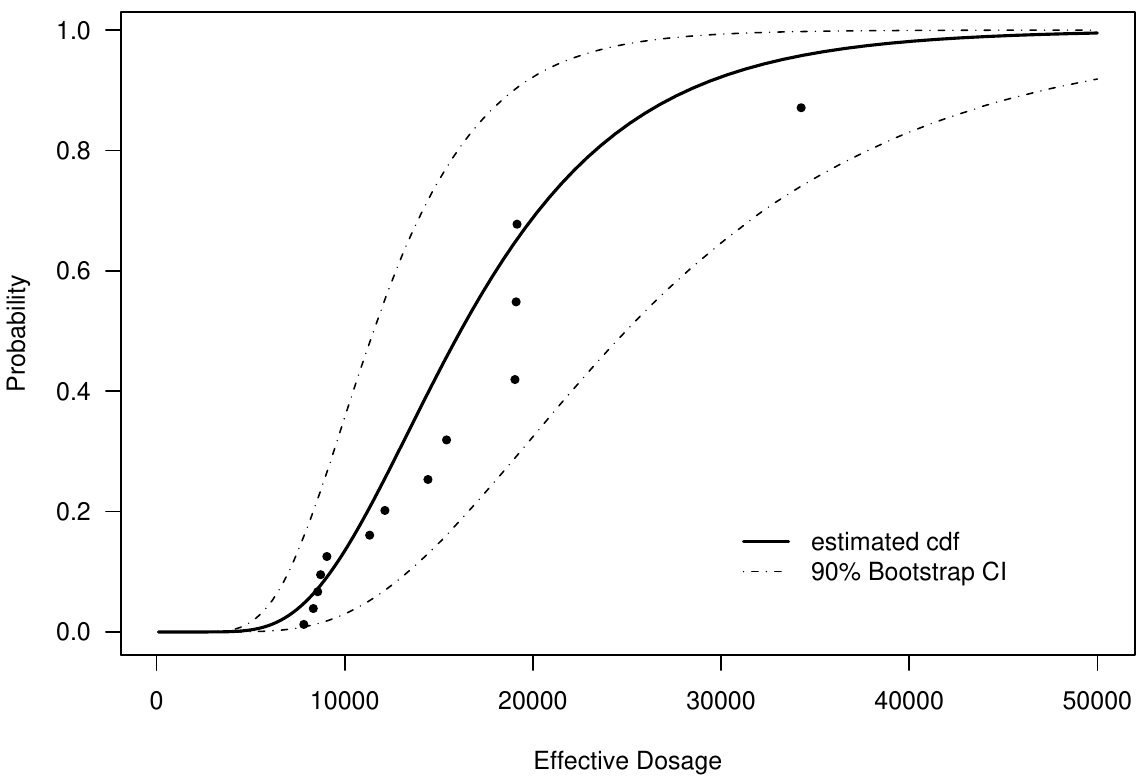}\\
(a) Laser Data & (b) Outdoor Weathering Data
\end{tabular}
\caption{Estimated $F(t)$ from the degradation model with pointwise 90\% CIs for the laser data (a) and outdoor weathering data (b). Dots show the Kaplan-Meier estimates. \emph{Figure reproduced from Meeker et al.~(2011) with permission from Wiley.}}
\label{fig:Ft.laser.weather.data}
\end{figure}

Following \citet{Meekeretal2014DegradationBookChapter}, we illustrate the analysis of RMDT data using the laser and outdoor weathering datasets. For the laser data, a linear degradation path of the form $\D(t) = \alpha + \beta t$ is used, where $\alpha$ is fixed (still needs to be estimated) and $\beta$ is a random variable that follows $\text{Normal}(\mu, \sigma)$. For this simple model, the CDF $F(t)$ has a closed-form expression:
\begin{align}\label{eqn:cdf.lognormal.rate}
F(t;\alpha,\mu,\sigma) = \pr[\D(t) \geq \thres]
= \Phi_{\NOR}\left\{ \frac{\log(t) - [\log(\thres - \alpha) - \mu]}{\sigma} \right\}.
\end{align}
Figure~\ref{fig:Ft.laser.weather.data}(a) displays the estimated $F(t)$ with pointwise 90\% CIs for the laser data, along with the Kaplan-Meier estimates~\citep{KaplanMeier1958}. The Kaplan-Meier estimate is a commonly used non-parametric estimate of the survival function.

The outdoor weathering degradation paths are nonlinear. We use the model in \eqref{eqn:log.logistic.form} for $\D(t)$, treating $\nu$ and $\kappa$ as random effects. The algorithms in \citet{Meekeretal2014DegradationBookChapter} are applied. Figure \ref{fig:Ft.laser.weather.data}(b) presents the estimated $F(t)$ with pointwise 90\% CIs, along with the Kaplan-Meier estimates. In this instance we see an estimated curve with a different shape and more uncertainty in some regions.

To incorporate accelerating variables, the practitioner must specify how they enter the model. For example, the Arrhenius relationship is commonly used when temperature is the accelerating variable. For other accelerating variables, functional forms such as the power law, inverse power, or log-linear relationships may be appropriate. Under the Arrhenius model, the transformed variable is defined as
$
x_i = 11605/(\temp_i + 273.15),
$
where $\temp_i$ is the temperature in degrees Celsius. This transformation converts the temperature to a Kelvin measurement.

\begin{figure}
\centering
\includegraphics[width=0.55\textwidth]{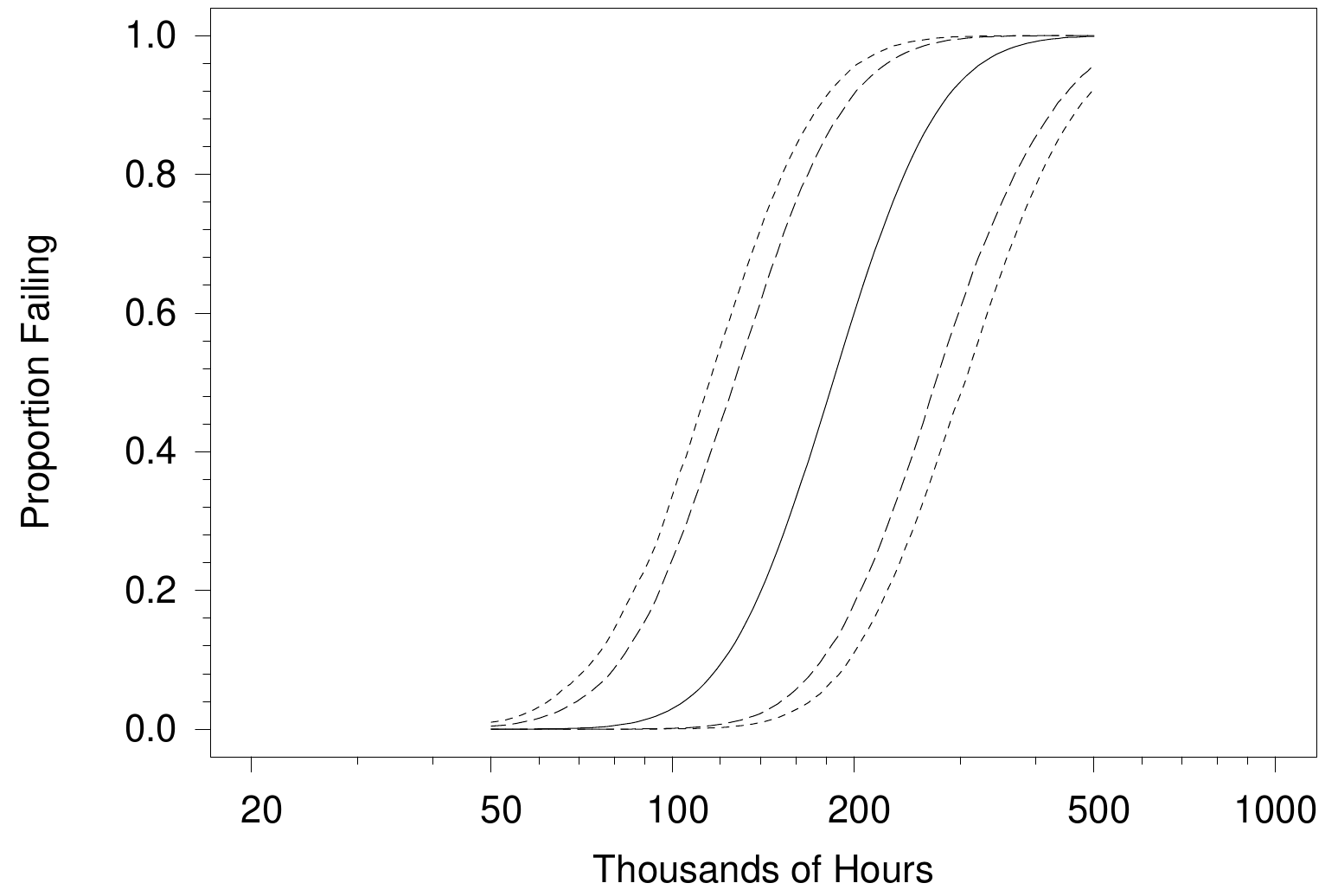}
\caption{Estimated CDF for Device B at 80 $\degreeC$ (solid line), with approximate 80\% and 90\% pointwise CIs (dotted lines). \emph{Figure reproduced from Meeker et al.~(2011) with permission from Wiley.}}
\label{fig:Ft.device.b.data}
\end{figure}

We use the Device B data to illustrate the maximum likelihood approach. Consider the following GPM:
\begin{align}
\D(t; x_i) = -\exp(\beta_2) \left(1 - \exp\left\{ -\exp(\beta_1) \exp[\alpha(x_0 - x_i)] t \right\} \right).
\end{align}
Here, $x_0 = 11605 / (195 + 273.15)$ (i.e., set $195\degreeC$ as the baseline temperature). We treat $\beta_1$ and $\beta_2$ as random effects, while the activation energy $\alpha$ is assumed to be fixed. Maximum likelihood estimation is used to obtain the parameter estimates. Figure~\ref{figure:device.b.data} shows the fitted degradation paths for all the test units in the Device B data. Figure~\ref{fig:Ft.device.b.data} presents the estimated failure-time distribution at 80$\degreeC$, under $\thres = -0.5$ dB.

\subsection{Bayesian Methods}\label{sec:Bayesian.method}

A Bayesian framework for the GPM based on RMDT data may also be used. The close connection between random effects models and Bayesian hierarchical priors allows us to conveniently leverage Bayesian inference for modeling the random effects.

\begin{figure}
\begin{tabular}{cc}
\includegraphics[width = 0.52\textwidth] {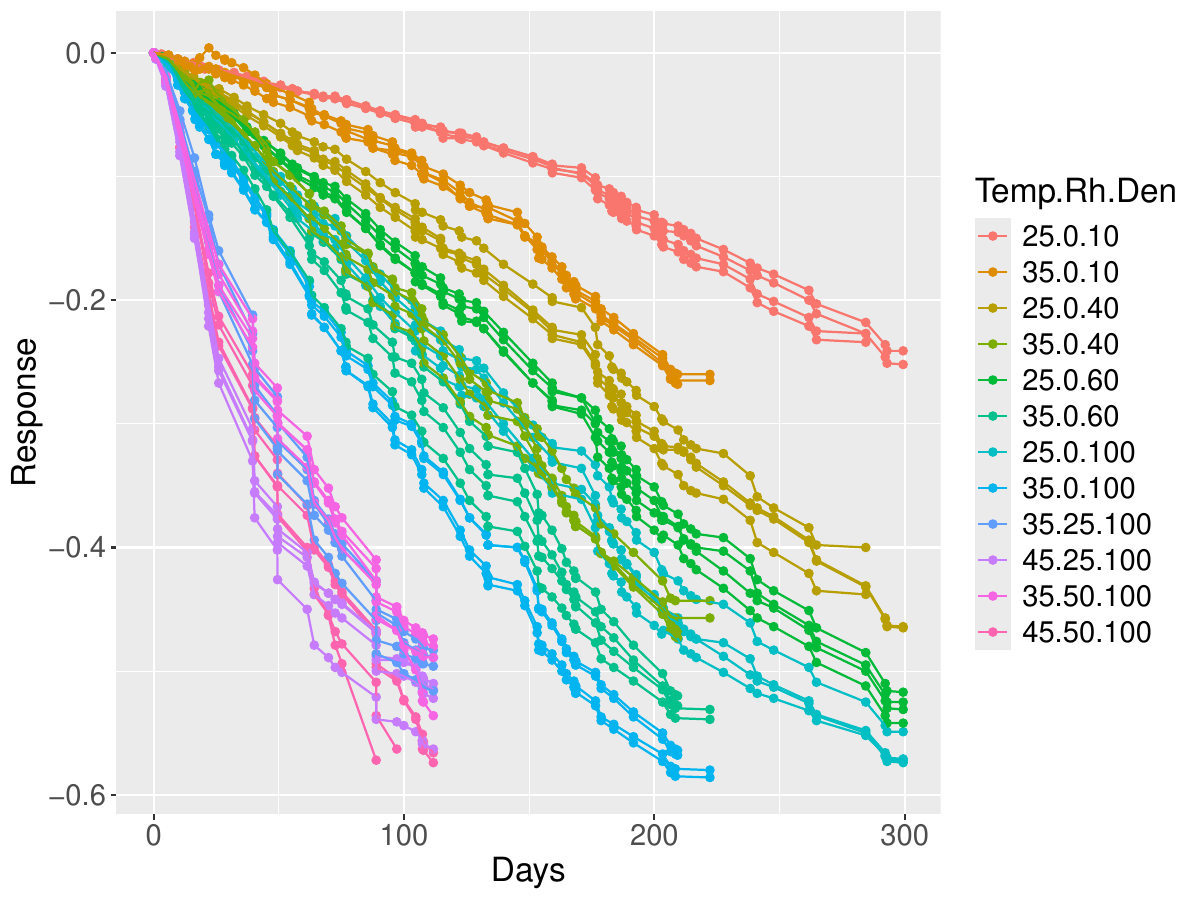}&
\includegraphics[width = 0.45\textwidth]{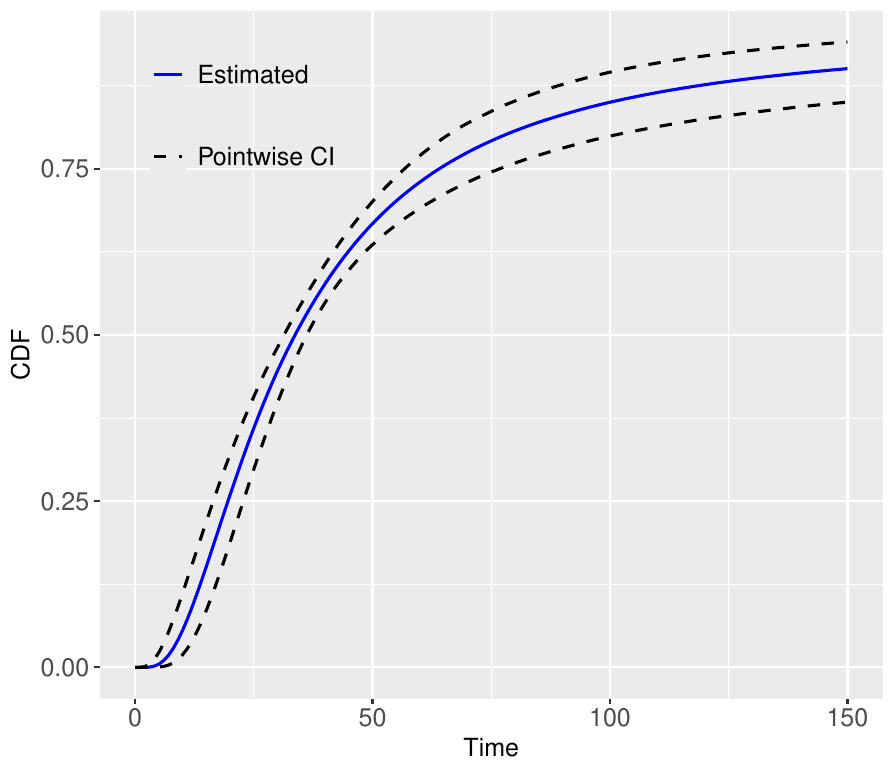} \\
(a) Degradation Data& (b) Estimated CDF \\
\end{tabular}
\caption{Visualization of the subset of the coating data  (a) and the estimated CDF  and pointwise  $95\%$ CIs (b).}\label{fig:coat.bayes.data.cdf}
\end{figure}

For illustration, we use the coating data from \cite{lu2021general}. In order to study the photodegradation process, scientists at the National Institute of Standards and Technology (NIST) conducted laboratory
tests at different levels of UV radiation and environmental conditions and measured
the chemical degradation of coating specimens every few days. The degradation of three different
types of chemical compounds were measured at three different wavelengths on 48 samples tested
under 12 test conditions at different combinations of UV intensity, temperature, and relative
humidity. For illustration purposes, we only take the subset of data measured at a wavelength of 1250
cm$^{-1}$. The explanatory variables are the UV intensity (Den), relative humidity (RH), and temperature (Temp).
Figure~\ref{fig:coat.bayes.data.cdf}(a) displays the observed degradation data at different levels of the three explanatory variables. In particular, UV intensity can take values $10 \%$, $40 \%$, $60 \%$, and $100\%$; relative humidity can take values $0 \%$, $25\%$, and $50 \%$; and temperature can take values  $25\degreeC$,  $35\degreeC$,and  $45\degreeC$. We apply the model in \eqref{eqn:log.logistic.form} with a re-parameterization for modeling convenience such that,
\begin{align}\label{eqn:ndpm}
\D_{i}(t_{ij}, \xvec_i, w_{i}; \etavec)&=\frac{\alpha\exp(w_{i})}
{1+\exp\left[-\dfrac{\log(t_{ij})-\mu_{i}(\xvec_i)}{\gamma}\right]}\,.
\end{align}
Note that the notation differed from that in \eqref{eqn:degradation.path.ran.eff}. Specifically, we use $w_i$ to represent the random effect for unit $i$, and denote the variance of the independent normally distributed random effects using $\sigma_w^2$. Additionally, $\mu_i$ is the location parameter of unit $i$, modeled as
\begin{align*}
\mu_i(\xvec_i) = \mu + \xvec_i'\betavec,
\end{align*}
where $\mu$ is the fixed intercept, $\xvec_i$ denote the transformed explanatory variables, using the same transformation form as in \citet{lu2021general}, and $\betavec$ includes the fixed regression coefficients. Further, $\gamma > 0$ is a scale parameter while $\alpha<0$ controls the asymptote when $w_i = 0$. The vector $\etavec$ in \eqref{eqn:ndpm} is defined as $\etavec = (\alpha, \mu, \gamma, \betavec')'$.

In the coating data analysis, we use flat or weakly informative priors for the parameters and their transformations. Specifically, we assign $\mu$, each element in $\betavec$, $\log(-\alpha)$, and $\log(\gamma)$ independent $\text{Normal}(0, 200)$ priors, and use inverse-gamma priors $\text{IG}(0.001, 0.001)$ for $\sigma_\varepsilon$ and $\sigma_w$. Recall that $\sigma_\varepsilon$ is the standard deviation of the error term in \eqref{eqn:degradation.path.ran.eff}.

Let $\thetavec = (\etavec', \sigmaeps, \sigma_w)'$. The joint posterior PDF of $\thetavec$ and $\wvec = (w_1,\ldots, w_n)'$ is expressed as
\begin{align}\label{eq:ceq1}
f(\thetavec, \wvec | \yvec) &\propto L(\thetavec| \yvec, \wvec) \times f(\wvec| \sigma_w) \times \pi(\thetavec) \\\nonumber
& \propto \prod_{i =1}^n   \left\{\prod_{k = 1}^{m} \frac{1}{\sigmaeps} \phi_{\NOR} \left[\frac{y_{ij} - \D_{i}(t_{ij}, \xvec_i, w_{i}; \etavec)}{\sigmaeps}\right] \times
{\frac{1}{\sigma_w}} \exp\left(-\frac{w_i^2}{2\sigma_w^2}\right)  \right\}  \times \pi(\thetavec),
\end{align}
where $\pi(\thetavec)$ denotes the joint prior distribution of $\thetavec$.

For inference, we use Markov chain Monte Carlo (MCMC) methods to draw samples of $\thetavec$,  along with the random effects $\wvec$, from the joint posterior distribution. In many Bayesian analyses it is not possible to explicitly solve for the posterior distributions. However, the product of the likelihood and the prior define a function which is proportional to the posterior. MCMC algorithms use the shape of the posterior distribution to produce samples from the target distribution. For more detailed discussions on Bayesian methods in reliability analysis, we refer readers to Chapter 10 of \citet{meeker2021statistical}, and to Chapter 21 of the same book for Bayesian approaches to degradation data analysis.

After convergence of the chains, MCMC samples can provide an approximation of the posterior distribution, and can be used in the estimation and inference of unknown parameters and functions of those parameters. In this analysis, the \texttt{RStan} package in R is used to implement the No-U-Turn Sampler \citep{hoffman2014no}. Posterior medians are used as point estimates for all parameters.

For reliability analysis, the CDF for model~\eqref{eqn:ndpm} is calculated as
\begin{align}\label{eqn:coating.cdf.Bayes}
F(t; \thetavec) &=\Pr \left[\D(t, \xvec, w) \leq \thres \right]=\Pr \left\{\log[-\D(t, \xvec, w)] \geq \log(-\thres) \right\},
\end{align}
where $\xvec$ and $w$ denote the vector of explanatory variables and random effect of a unit, respectively. Note that here the path is decreasing. This CDF is conditional on unknown parameters $\thetavec$. Uncertainty in these unknown parameters is captured in the posterior distribution. We take $F_{w}(\cdot)$ to be the CDF of $w$ and define $$\xi_t = \log(\thres/\alpha)+\log \left\{1 + \exp \left[  - \frac{\log(t) - (\mu + \xvec' \betavec)}{\gamma} \right]\right\}.$$
Note that $\xi_t$ is the argument such that $F(t;\thetavec) = 1-F_{w}(\xi_t|\thetavec)$, which can be obtained by rearranging terms in $\log[-\D(t, \xvec, w)] \geq \log(-\thres)$.

For each draw of the parameters, the CDF in \eqref{eqn:coating.cdf.Bayes} is calculated, producing a draw of the CDF. The point estimates (taking $\thetavec$ to be the posterior mean or median) and credible intervals can then be obtained using draws of the CDF. Figure~\ref{fig:coat.bayes.data.cdf}(b) shows the estimated CDF in \eqref{eqn:coating.cdf.Bayes} and the associated pointwise $95\%$ credible intervals at UV intensity of $90 \%$, RH of $30 \%$, and temperature of $40\degreeC$, with $\thres=-0.4$.

\subsection{Remaining Useful Life}

The prediction of the remaining useful life (RUL) is crucial in applications such as system health monitoring and condition-based maintenance, two common scenarios where degradation modeling is employed. RUL refers to the amount of time that a system can still be used effectively, and estimating the distribution of RUL is often useful to users of the system.
The failure time, for unit $i$, can be expressed as a function $T(\thres, w_i)$, given the random effect $w_i$ and failure threshold $\thres$.
Given the observed degradation path for unit $i$ up to time $t_{0i}$, the CDF of the RUL is,
$$
\rho_i(s; \thetavec) =  \pr\left[ T(\thres, w_i) \leq t_{0i} + s \,\middle|\, T > t_{0i} \right] , \quad s > 0.
$$
Here, $\rho_i(s; \thetavec)$ is the conditional probability that failure occurs within $s$ time units after $t_{0i}$, with respect to the conditional distribution $w_i | \yvec$. While there may be interest in prediction uncertainty for RUL, practitioners are often interested in the CDF for the distribution of the RUL, with uncertainty in the CDF introduced by the hyperparameters. In our examples, we typically focus on the CDF of the RUL distribution.

It is helpful to illustrate the Bayesian framework with an example. We continue the analysis of the coating data from Section~\ref{sec:Bayesian.method}. In this case, the RUL distribution can be evaluated as,
\begin{align*}
\rho_i(s; \thetavec) & =\int_{w_i} \pr\left( \D(t_{0i} + s, \thetavec, w_i) \leq \thres | \D(t_{0i}, \thetavec, w_i) \geq \thres, \thetavec, w_i \right) f( w_i |\yvec) dw_i,
\end{align*}
where $f( w_i | \yvec)$ denotes the marginal posterior distribution of $w_i$. MCMC methods are typically used to perform this computation. Since the MCMC samples of $w_i$ are draws from the conditional distribution of $w_i  | \yvec$, they are used to calculate the draws of $\pr\left( \D(t_{0i} + s, \thetavec, w_i) \leq \thres | \D(t_{0i}, \thetavec, w_i) \geq \thres, \thetavec, w_i \right)$.  The MCMC samples of  $\thetavec$ are then substituted into the expression to incorporate the uncertainty from $\thetavec$, and the point estimates and credible intervals are obtained based on the draws. The Bayesian framework offers a versatile approach for updating inference as new degradation data become available.

\begin{figure}
\centering
\includegraphics[width = 0.45\textwidth]{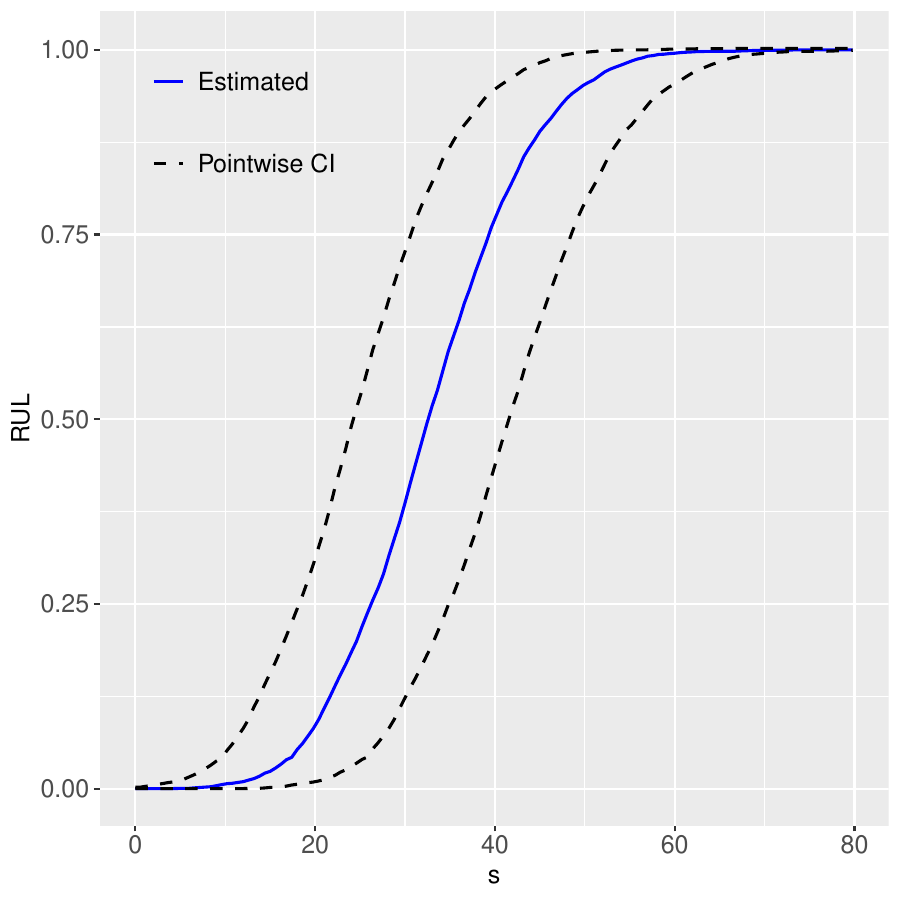}
\caption{The estimated CDF of RUL and associated pointwise $95\%$ CI for a unit tested at  UV intensity $40\%$, relative humidity $50 \%$, and temperature $35\degreeC$ and has not failed at $t_0 = 150$.}\label{fig:coat.bayes.RUL}
\end{figure}

Using MCMC samples, the estimated CDF of RUL and corresponding credible intervals for a unit that has not yet failed at $t_{0} = 150$ is calculated. The unit is tested at UV intensity $40 \%$, relative humidity $50\%$, and temperature $35\degreeC$. Figure~\ref{fig:coat.bayes.RUL} illustrates the estimated CDF of RUL and associated pointwise $95\%$ credible interval for this unit.

\subsection{General Path Model for ADDT Data}\label{sec:ADDT.inf}

A commonly used model for ADDT data is
\begin{align}\label{eqn:ADDT.gen.model}
y_{ijk} = \D(t_{ij}; x_j, \thetavec) + \epsilon_{ijk},
\end{align}
where notation was introduced in Section~\ref{sec:ADDT.data}. Note that this model still belongs to the general class of GPMs. However, for ADDT data, only the population-level degradation path can be modeled, as repeated measurements on the same unit are not available.

For the Adhesive Bond B data, \cite{Xieetal2018} use the parametric model
\begin{align} \label{eqn: AdhesiveBondB parametric model}
y_{ijk}=\beta_{0}+\beta_{1}\exp(\beta_{2} x_{i} ) \tau_{ij} + \varepsilon_{ijk},
\end{align}
where $y_{ijk}$ represents the strength of Adhesive Bond B, measured in log Newtons; $\tau_{ij} = \sqrt{t_{ij}}$ is the transformed time variable; and $x_i = -11605 / (\temp_i + 273.15)$ is the transformed temperature. The error term is assumed to follow $\epsilon_{ijk} \sim \text{Normal}(0, \sigma)$, with correlation $\text{Corr}(\epsilon_{ijk}, \epsilon_{ijk'}) = \rho$ for $k \neq k'$ to account for the dependence among measurements taken under the same time and temperature conditions.

Inference for parametric models is based on maximum likelihood estimation; see \citet{King2016} for details. Briefly, the likelihood function is
\begin{align}\label{eqn:likelihood}
L(\thetavec) = \prod_{i,j}(2\pi)^{-n_{ij}/2} |\Sigma_{ij}|^{-1/2} \exp\left\{-\tfrac{1}{2}[\yvec_{ij} - \muvec(t_j, x_i)]'\Sigma_{ij}^{-1}[\yvec_{ij} - \muvec(t_j, x_i)]\right\},
\end{align}
where $\yvec_{ij} = (y_{ij1}, \dots, y_{ijn_{ij}})'$ is the vector of degradation measurements. The mean vector $\muvec(t_j, x_i)$ has all entries equal to $\D(t_j, x_i)$, and $\Sigma_{ij}$ is an $n_{ij} \times n_{ij}$ covariance matrix with diagonal entries $\sigma^2$ and off-diagonal entries $\rho \sigma^2$. Here, $\thetavechat$ maximizes this likelihood.

\citet{Xieetal2018} also propose a semiparametric model, which is represented as
\begin{align}\label{eqn: semi-parametric degradation model}
\D(t_{ij}, x_i; \thetavec)= g\left[\eta_{i}(t_{ij};\beta);\gammavec\right],
\quad \eta_{i}(t; \beta) =\frac{t}{\exp{(\beta s_i)}},\quad s_i =x_{\max}-x_i.
\end{align}
Here the $g(\cdot)$ function is modeled by monotonic B-splines. Note that semiparametric models offer high flexibility in representing the degradation path while retaining interpretable effects of explanatory variables. For the estimation and inference procedures, we refer to \citet{Xieetal2018}.

\begin{figure}
\begin{center}
\begin{tabular}{cc}
\includegraphics[width=0.48\textwidth]{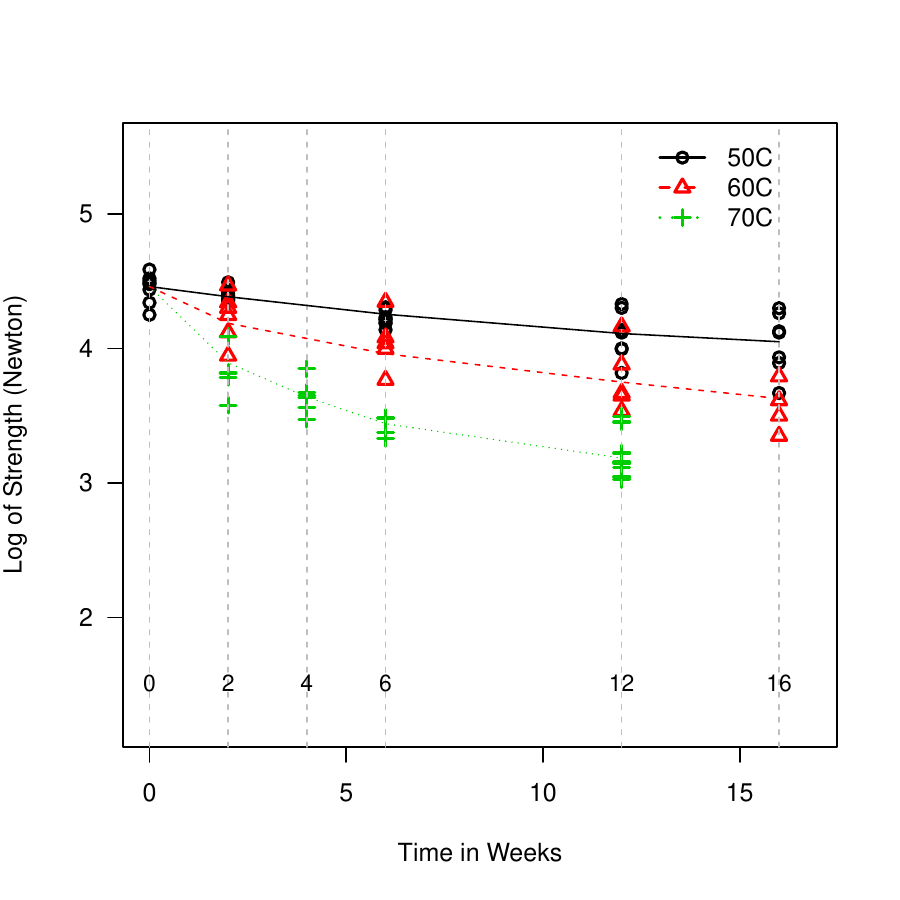}&
\includegraphics[width=0.48\textwidth]{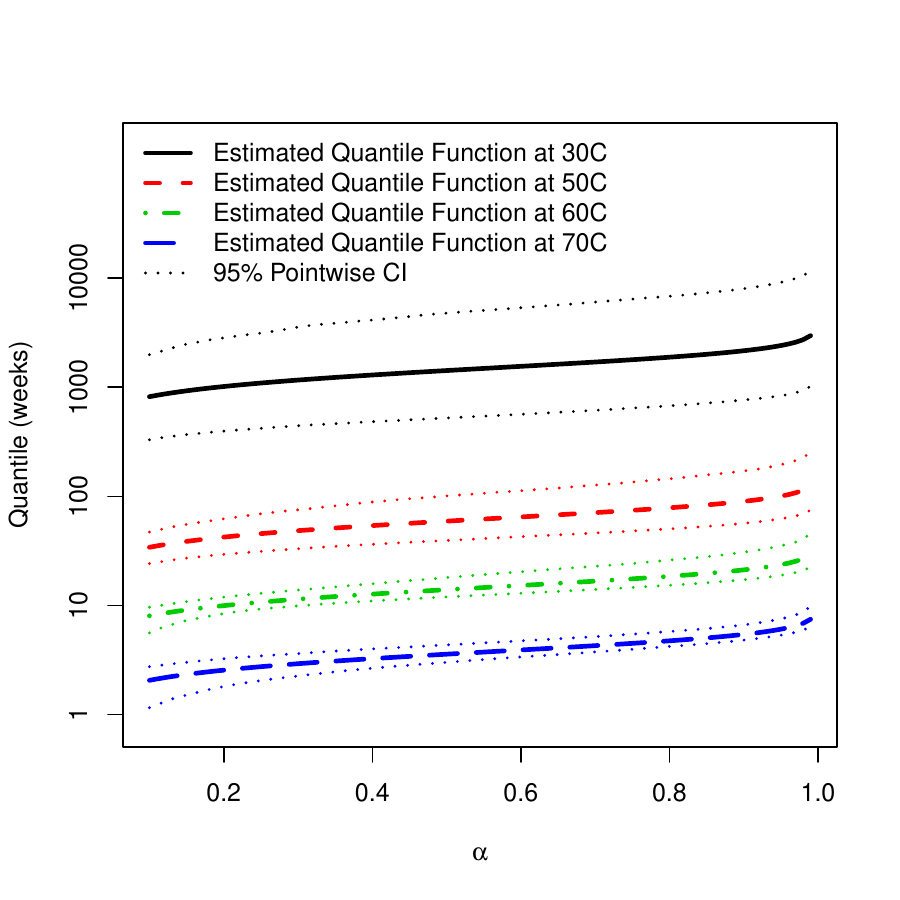}\\
(a) Semiparametric Fit & (b) Estimated Quantiles
\end{tabular}
\end{center}
\caption{Visualization of the fitted semiparametric degradation paths (a) and the estimated failure time quantile functions (b), based on the Adhesive Bond B data. \emph{Figure reproduced from Xie et al.~(2018) with permission from Taylor and Francis.}} \label{fig:AdhesiveBondB.data.sm.fit.qf}
\end{figure}

The failure time distribution from the ADDT model can be derived by noting that, for monotonic decreasing paths, the event $T \leq t$ is equivalent to the degradation measurement $y_t \leq \thres$ at time $t$. For the semiparametric model, the CDF of the failure time, $F(t)$, is
$$
F(t) = \Pr(y_t \leq \thres) = \Phi_{\nor} \left( \frac{\thres - g\left[ t/\exp(\beta s); \gammavec \right]}{\sigma} \right), \quad t \geq 0.
$$
The $q$ quantile of the failure time distribution can be obtained by inverting the CDF: $t_q = F^{-1}(q)$. Figure~\ref{fig:AdhesiveBondB.data.sm.fit.qf} illustrates the use of the semiparametric model with the Adhesive Bond B data. The fitted degradation paths appear on the left and the estimated quantiles appear on the right. We note a general decrease in the value of each quantile as temperature increases.

\subsection{Stochastic Process Models}

We briefly outline the parameter estimation procedures for SP models. The likelihood function is given by
\begin{align}\label{eqn:sto.lik}
L(\thetavec \mid \text{Data}) = \prod_{i=1}^n \prod_{j=0}^{m_i - 1}
f[\Delta y_i(t_{ij}); \Delta\mu(t_{ij}),\thetavec],
\end{align}
where $\thetavec$ represents the vector of unknown parameters. The increment $\Delta y_i(t_{ij}) = y_i(t_{i,j+1}) - y_i(t_{ij})$ is the observed degradation change between two consecutive time points, and $\Delta\mu(t_{ij}) = \mu(t_{i,j+1}; \thetavec) - \mu(t_{ij}; \thetavec)$ is the corresponding change in the mean function. Here, $t_{i0} = 0$ and $m_i$ is the number of observations for unit $i$. The PDF $f(\cdot)$ is determined by the probability distribution associated with the specific SP (e.g., Wiener, gamma, or inverse Gaussian).

A parametric specification of the mean trend function $\mu(t)$ is required for model fitting. A commonly used form is the power-law function:
$
\mu(t) = \mu(t; \thetavec) = \left(t / \alpha_2\right)^{\alpha_1}.
$
Here $\alpha_2$ is a scaling in time while $\alpha_1$ is a growth rate. The maximum likelihood estimate of $\thetavec$ is obtained by maximizing the likelihood function given in \eqref{eqn:sto.lik}.

\begin{figure}
\begin{center}
\includegraphics[width=.55\textwidth]{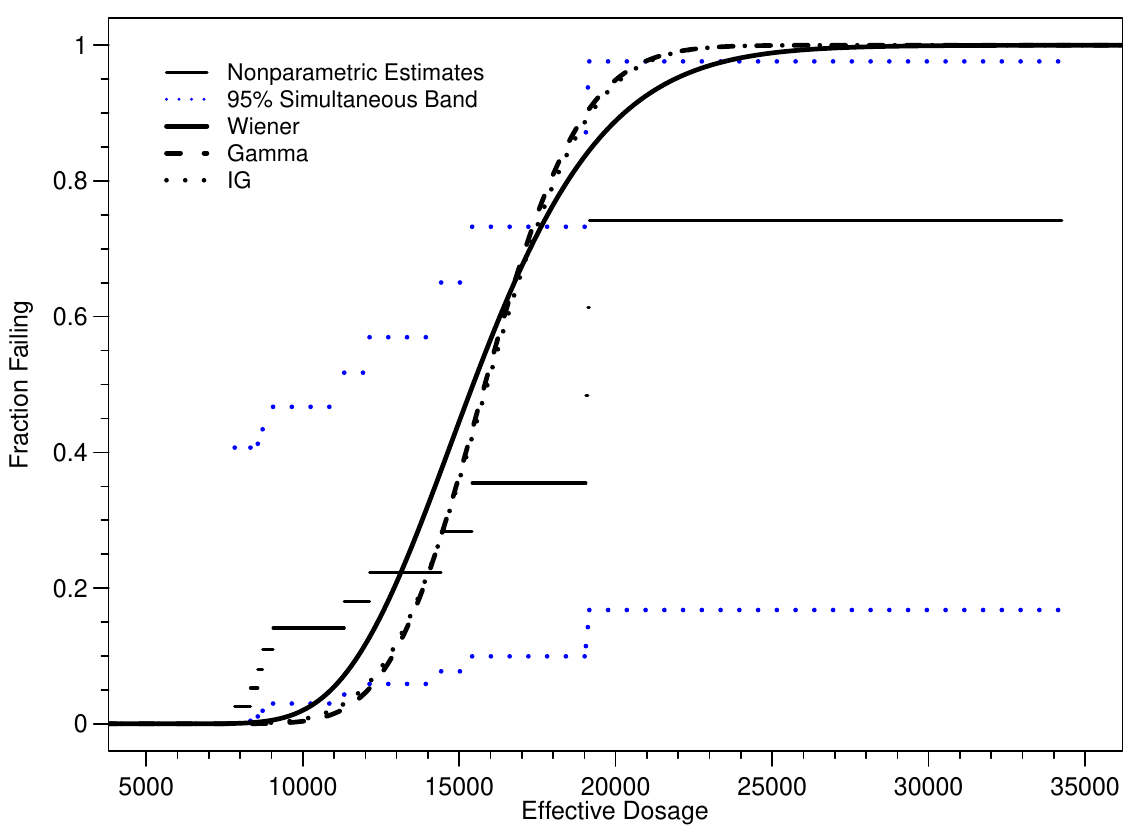}
\end{center}
\caption{Estimated CDFs from three stochastic models and the Kaplan-Meier estimate with approximate 95\% SCBs. \emph{Figure reproduced from Meeker et al.~(2011) with permission from Wiley.} }\label{fig:weathering.dat.sto.fit}
\end{figure}

As an illustration, we apply stochastic models with the power-law trend function from \citet{Meekeretal2014DegradationBookChapter}. Figure~\ref{fig:weathering.dat.sto.fit} displays the estimated CDFs from the three SP models, along with the Kaplan-Meier estimate and its 95\% simultaneous confidence bands (SCBs). The nonparametric confidence bands were constructed using the method of \citet{Nair1984}. Based on the plots, there is no clear evidence favoring one stochastic process model over the others.

\section{Case Studies of Degradation Models}\label{sec:applications}

In addition to the examples provided in Section~\ref{sec:inference.procedures}, which focus on electronics, coatings, and devices, degradation models have many other applications. Section~\ref{sec:coating} presents an additional example from the coating industry. Section~\ref{sec:polymeric.materials} discusses applications in polymeric materials. Section~\ref{sec:metal.fatigue} presents a brief overview of the analysis of fatigue crack growth phenomena, while Section~\ref{sec:civial.eng} provides examples from civil infrastructures.

\subsection{Coatings}\label{sec:coating}
Organic coatings are widely used in various applications, and their degradation is primarily driven by UV radiation, along with temperature and humidity. There is a substantial body of literature on modeling coating degradation. Here, we review a case study from \citet{HongDuanetal2015}, which models degradation using dynamic independent variables as described in Section~\ref{sec:RMDT.data}. The units saw random start times between 161 and 190 days. For degradation data with dynamic independent variables, the modeling approach in \citet{HongDuanetal2015} is
\begin{align}
y_i(t) = \D[t; x_{il}(t)] + R(t; w_i) + \epsilon(t),
\end{align}
where $y_i(t)$ denotes the degradation measurement for unit $i$ at time $t$. The mean structure is given by $\D[t; x_{il}(t)]$, which depends on time-varying independent variables indexed by $l = 1, 2, \dots, p$. The error structure consists of two components: $R(t; w_i)$ represents a functional random effect for unit $i$, relying on scalar random effect $w_i$, and $\epsilon(t)$ captures the independent noise at time $t$.

One way to model the mean degradation path is:
\[ \D[t; x(t)] = \beta_0 + \sum_{l=1}^p \int_0^t f_l[x_l(\tau); \beta_l]d\tau. \]
where $f_l(\cdot;\cdot)$ is the functional effect for independent variable $l$. The full term,
$\int_0^t f_l[x_l(\tau); \beta_l]d\tau,$
is the cumulative effect of independent variable $l$ at time $t$. Nonparametric spline methods are often used to model $f_l$ due to their flexibility and ability to incorporate constraints.

To predict failure time and reliability for a random unit, the independent variable process $\Xvec(\infty)$ and random effect $w$ are treated as random. Thus, the CDF of the failure time, defined as $T$, is given by:
\begin{align}\label{eqn:Ft}
	F(t;\thetavec)=\pr\left\{T[\D_f,\Xvec(\infty),\wvec]\leq t\right\},\quad t>0.
\end{align}
Computing details are described in \cite{HongDuanetal2015}.

\begin{figure}
\begin{center}
\includegraphics[width=0.45\textwidth]{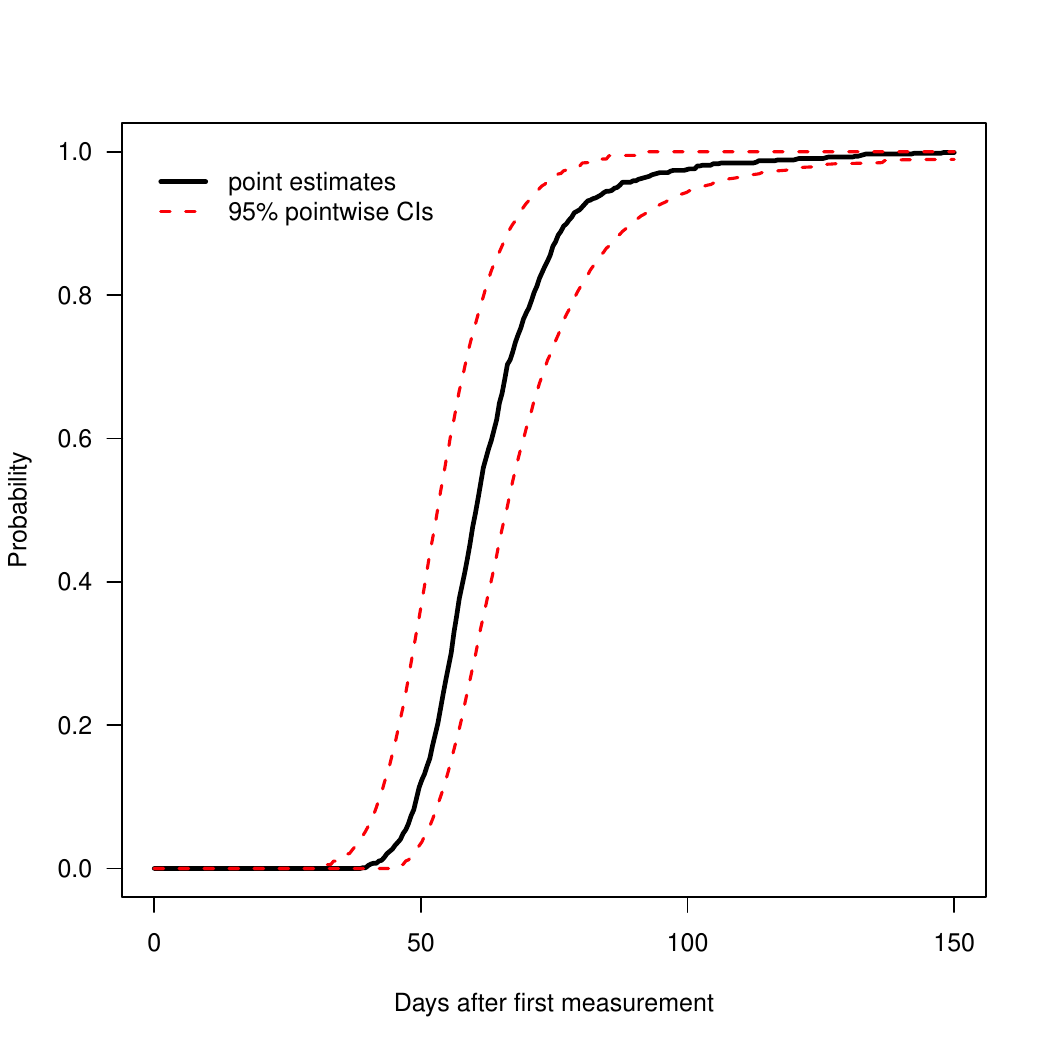}
\caption{Estimated CDF and 95\% pointwise CIs for a population with random start times between 161 and 190 days. \emph{Figure reproduced from Hong et al.~(2015) with permission from Taylor and Francis}.}\label{fig:rand.ind}
\end{center}
\end{figure}

Figure~\ref{fig:rand.ind} illustrates the estimated $F(t; \thetavec)$ and its 95\% pointwise CIs for the coating degradation data. For units with random start times between 161 and 190 days, most failures occur between 50 and 150 days in service. \citet{Duanetal2017} provides additional modeling work on coating degradation, in which the authors use a physically motivated approach to incorporate information from independent variables. The authors focus on a cumulative damage model with random effects for analyzing accelerated test data. Further effort in modeling coating degradation accelerated laboratory test data is provided by \citet{lu2021general}. In that work, the authors focus on modeling multiple correlated degradation characteristics and present an expectation-maximization based algorithm for fitting the proposed model.

\subsection{Polymeric Materials}\label{sec:polymeric.materials}

Polymeric materials, or polymers, are a large class of materials used in products ranging from everyday items such as clothing and food packaging to specialty medical devices. With benefits such as strength, flexibility, heat resistance, and lightweightness, polymers are becoming increasingly more common in multiple industrial applications. For example, polymers often replace metal in aerospace components as the polymeric materials provide better performance in extreme conditions at a lower weight.

When selecting materials for an application, material scientists and engineers frequently employ ADDT to estimate the thermal index (TI) of a polymer. The TI provides a measure of the polymeric material's long-term performance. \cite{UL746B} describes the TI as ``an indication of the material's ability to retain a particular property (physical, electrical, etc.) when exposed to elevated temperatures for an extended period of time'' (p.\ 7). For example, a polymer with a TI of 200$\degreeC$ can be expected to maintain a property, such as strength, at a temperature of 200$\degreeC$ for 100,000 hours. \cite{Xieetal2017ThermalIndex} present a detailed approach for estimating the TI using ADDT data. Based on the GPM model for ADDT data in (\ref{eq:ceq1}), temperature is treated as the accelerating variable, following the Arrhenius relationship. The mean time to failure $m(x)$ is solved from $\D[m(x),x]=\D_0$. That  is, $m(x)=\D^{-1} (\D_0,x)$. For a specified target time to failure $t_d$, the corresponding temperature TI is determined by solving for $x_d$ in
\[ m(x_d)=t_d. \]
In practice, $t_d=100{,}000$ hours is often used. Given
\[x_d=m^{-1} (t_d )=1/(\ti+273.16),\]
the corresponding temperature TI is
\begin{align}\label{eqn:ti}
\ti=1/(m^{-1} (t_d ) )-273.16.
\end{align}

The temperature TI in (\ref{eqn:ti}) defines the material's thermal index. Note that the constant 273.16 is the conversion factor between the Celsius and Kelvin temperature scales.
Estimating TI involves first determining the degradation path from the ADDT data, as outlined in Section \ref{sec:ADDT.inf}. \cite{Xieetal2017ThermalIndex} provide a comprehensive comparison of degradation modeling and TI estimation techniques using the Adhesive Bond B data. Figure 13 shows the fitted temperature-time curves and estimated TIs for Adhesive Bond B using the traditional (TM), parametric (PM), and semiparametric (SPM) methods. The PM and SPM methods, which \cite{Xieetal2017ThermalIndex} recommend over TM, yield similar TI estimates of 33$\degreeC$ and 34$\degreeC$, respectively.

\begin{figure}
\begin{center}
\includegraphics[width=.45\textwidth]{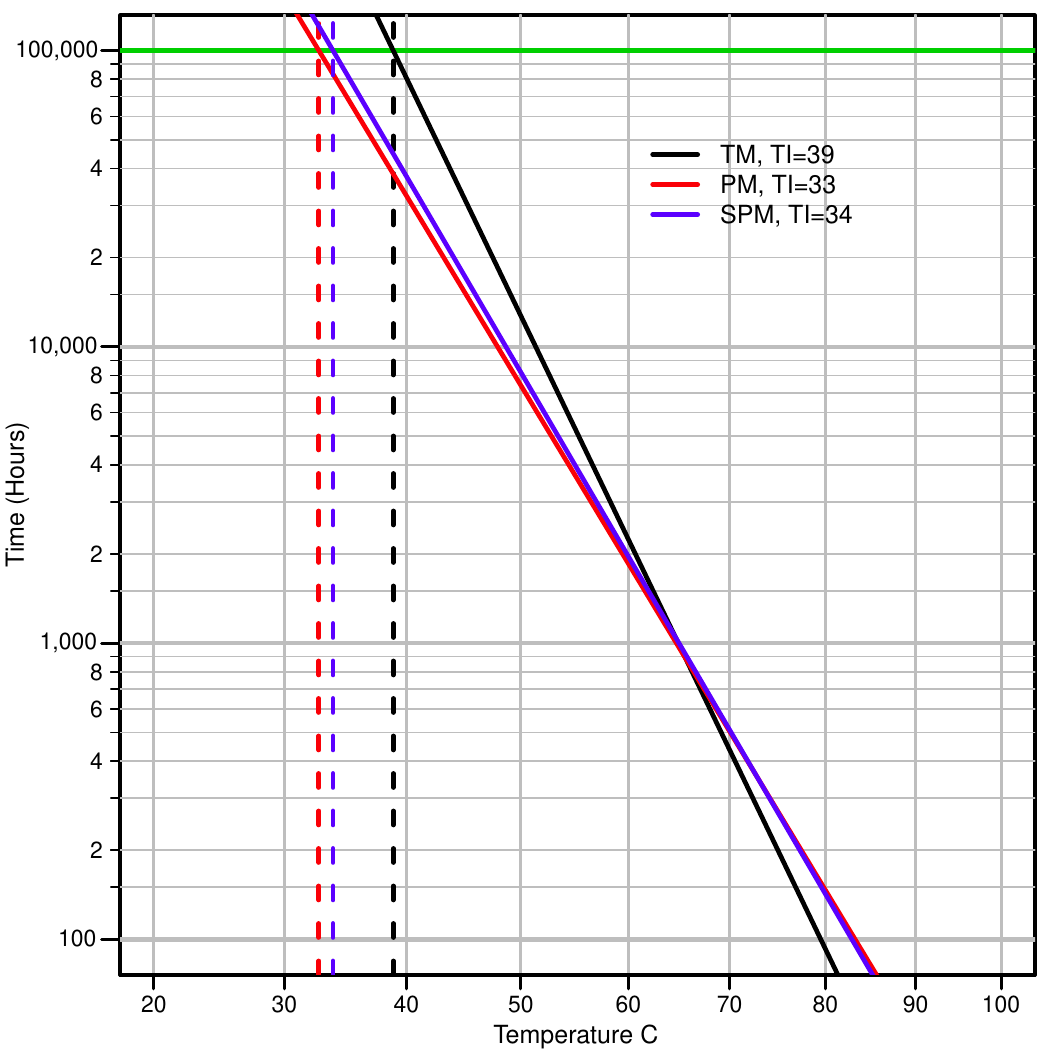}
\end{center}
\caption{Fitted temperature-time curves and estimated TIs for Adhesive Bond B using the TM, PM, and SPM. \emph{Figure reproduced from Xie et al.~(2017) with permission from Springer}.
}\label{fig:adhesive.bond.b.TI}
\end{figure}

\subsection{Metal Fatigue Crack Growth}\label{sec:metal.fatigue}
Metal fatigue crack growth is a pervasive and challenging problem in structural engineering, characterized by the progressive degradation and eventual fracture of materials subjected to cyclic loading. Crack growth phenomenon often occurs silently, initiating as microscopic cracks that gradually grow until catastrophic failure ensues, which can occur without significant prior deformation.

Historical incidents highlight the seriousness of fatigue-induced crack propagation, emphasizing the need for standard inspection protocols, growth tracking, and accurate predictive methods.  Notable catastrophes attributed to metal fatigue crack growth include the mid-air breakups of the De Havilland Comet jets in the early 1950s, where cracks initiated in high stress cut-outs,
such as windows, due to cyclic cabin pressurization.  These tragic failures led to explosive decompression and loss of life \citep{withey1997fatigue}.
More recently, the catastrophic Eschede train derailment in Germany in 1998 illustrated how undetected fatigue cracks in critical components like wheels could cause sudden structural failures, resulting in significant fatalities \citep{esslinger2004railway}. These high-profile cases show that fatigue crack growth poses real-world safety risks beyond academic interest.

\begin{figure}
\centering
\begin{tabular}{cc}
\includegraphics[width = 0.48\textwidth]{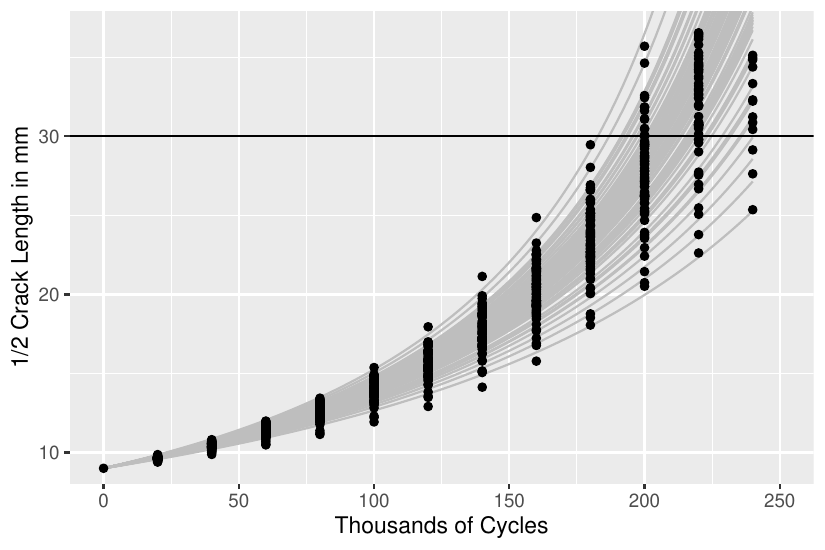}&
\includegraphics[width = 0.48\textwidth]{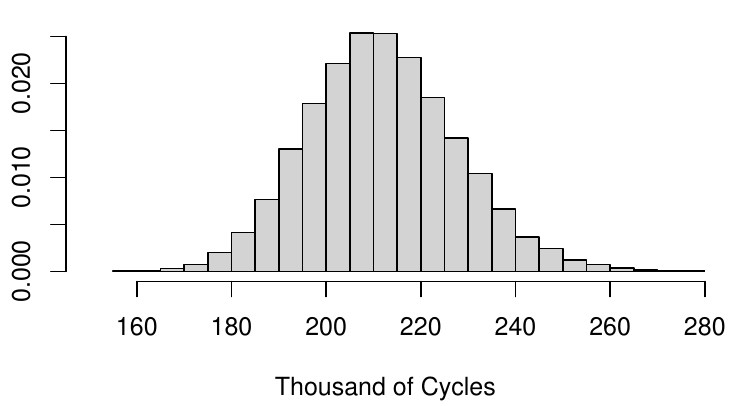}\\
(a) Crack Length & (b) Estimated Cycles
\end{tabular}
\caption{The fitted model with the observed data (a), and a histogram of the estimated number of cycles until the 1/2 crack length of a random specimen exceeds 30mm~(b). }\label{fig:virklerFittedPaths.hist} 
\end{figure}

Metal fatigue crack growth in traditional materials such as steel and aluminum cannot self-repair.  Thus subsequent measurements must be at least equal to or greater than previous measurements.  Additionally, as a general rule, crack growth accelerates as the length increases; as illustrated in Figure~\ref{fig:virklerFittedPaths.hist}(a).  These typical characteristics can help determine an appropriate model and can possibly inform prior distributions for parameters.

We provide an introductory example of metal fatigue crack growth using repeated measures data captured from Figure 3 in \cite{kozin1981critical}.  The original data were 1/2 crack lengths in a central slit aluminum tension test specimen; all of which start with a length of 9mm.  We will consider a derivative of the data which can be accessed at \url{https://github.com/WarrRich/Virkler-Data}, along with details of how the data were obtained, and R code for this example.

The 1/2 crack length measurements are displayed as dots in Figure~\ref{fig:virklerFittedPaths.hist}(a).  We use the Paris-rule model which was introduced earlier in \eqref{eqn:paris.rule}.  We implement a random effects model which allows for each test specimen to have its own curve and allows for some measurement error about that curve. This is accomplished using a Bayesian hierarchical model.
The model produces an estimated population of 1/2 crack length paths from these aluminum test specimens. We are interested in estimating the distribution (in number of cycles) in which the 1/2 crack length exceeds a threshold of 30mm.

In our model we assume $x$ is a fixed constant stress of 1 for all crack length paths and that the deviation from the crack-length paths for each observation are independent and identically distributed (iid) $\text{Normal}(0,\sigma_{\varepsilon})$. Let $t_{ij}$ represent the number of thousands of cycles for the $j$th measurement on specimen $i$.  Additionally, let $\theta_{1i}$ and $\theta_{2i}$ be the item specific parameters of the $i$th specimen.  We expand the notation of \eqref{eqn:paris.rule} and let $\mathcal{D}(t_{ij},\theta_{1i}, \theta_{2i})$ be a function of the number of cycles and the primary model parameters for specimen $i$.  The model is:
\[y_{ij} \mid \theta_{1i}, \theta_{2i}, \sigma_{\varepsilon} \stackrel{\mbox{\scriptsize iid}}{\sim} \text{Normal}\!\left( \mathcal{D}(t_{ij},\theta_{1i}, \theta_{2i}), \sigma_{\varepsilon} \right) .\]

We point out a few items that help formulate the prior distribution.  We know what reasonable values are for the specimen specific parameters $\theta_{1i}$ and $\theta_{2i}$.  For all $i$, we constrain $\theta_{1i} > 0$, otherwise it would produce a decreasing path, additionally, we observe that it should be fairly small, to produce the curves similar to those in Figure \ref{fig:virklerFittedPaths.hist}(a).  Next, from \eqref{eqn:paris.rule} we observe that when $\theta_2 = 2$ the crack growth rate is approximately an exponential growth rate.  Thus we center our prior of $\theta_2$ on 2.  Using this information we formulate the hierarchical prior as:
\begin{align} \label{eq:metalModel}
  \begin{split}
    \begin{bmatrix}
    \log(\theta_{1i}) \\
    \theta_{2i}
    \end{bmatrix} & \stackrel{\mbox{\scriptsize iid}}{\sim}  \,
    \N\left[
    \begin{pmatrix}
    \mu_{\theta_{1}} \\
    \mu_{\theta_{2}}
    \end{pmatrix},
    \begin{pmatrix}
    \sigma_{\theta_{1}}^2 & \sigma_{\theta}\\
    \sigma_{\theta} & \sigma_{\theta_{2}}^2
    \end{pmatrix}
    \right],
  \end{split}
\end{align}
$\mu_{\theta_{1}} \sim \text{Normal} \left(-9,1\right)$, $\mu_{\theta_{2}} \sim \text{Normal} \left(2,\sqrt{10}\right)$, $\sigma_{\theta_{1}} \sim \text{Exp}(1)$,
$\sigma_{\theta_{2}} \sim \text{Exp}(1)$, $\rho \sim \text{Unif}(-1,1)$, and $\sigma_{\varepsilon} \sim \text{Exp}(1)$, where $\sigma_{\theta} =\rho \, \sigma_{\theta_{1}} \sigma_{\theta_{2}}$.  Some choices for the parameters for $\mu_{\theta_{1}}$ and $\mu_{\theta_{2}}$ were selected by testing different values in \eqref{eqn:paris.rule} to roughly mimic the data.  The other hyperpriors were selected to be fairly diffuse around reasonable values of the parameters.

Posterior samples are obtained using MCMC, and the posterior means are shown as lines in Figure~\ref{fig:virklerFittedPaths.hist}(a). The model fits the data reasonably well. A primary goal of the analysis is to estimate the number of cycles until the 1/2 crack length exceeds 30mm. Figure~\ref{fig:virklerFittedPaths.hist}(b) shows a histogram of the estimated
number of cycles until the 1/2 crack length exceeds 30mm.

%

\subsection{Civil Infrastructures}\label{sec:civial.eng}
Aging civil infrastructures, including transportation and water systems, are increasingly underperforming and becoming structurally deficient, requiring urgent attention. For example, 43\% of major US roads are in poor or mediocre condition, costing motorists an estimated USD 130 billion annually \citep{mingy1}. Effective maintenance planning and improved infrastructure durability rely on accurate degradation modeling, which must account for the complex and uncertain influences of materials, design, usage, environment, and maintenance.

\begin{figure}
	\centering
	\includegraphics[width=0.7\textwidth]{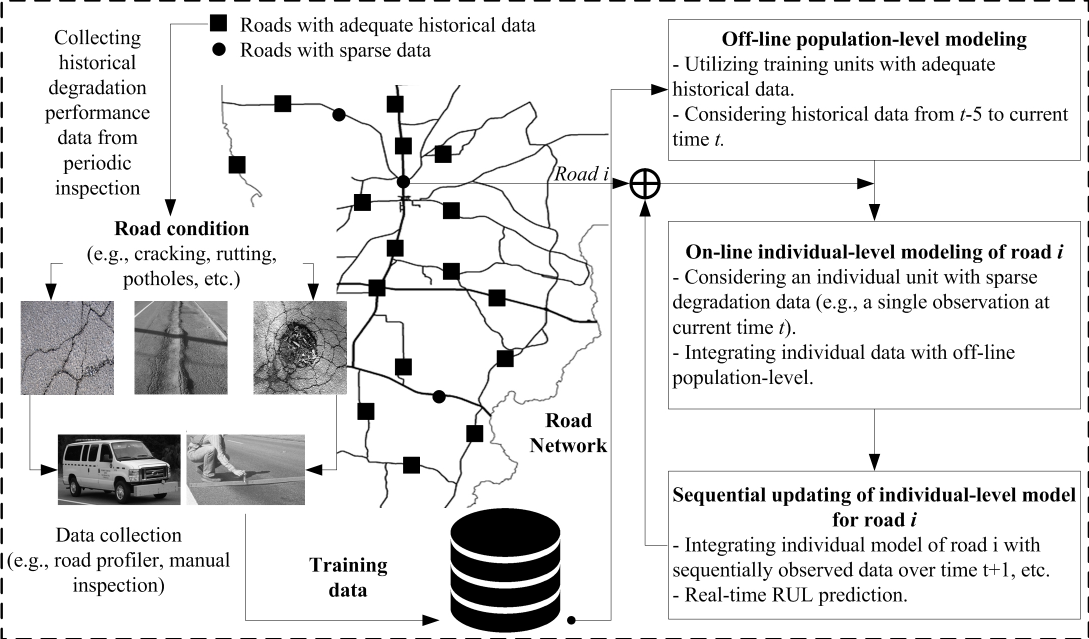}
	\caption{Illustration of degradation modeling framework for roads. \emph{Figure reproduced from Nguyen et al.~(2023) with permission from Wiley}. }\label{fig:civ1}
\end{figure}

Here, we revisit a case study by \citet{nguyen2023bayesian}, which propose a heterogeneous degradation modeling framework under a nonparametric Bayesian approach to predict the RUL of individual road sections. Figure~\ref{fig:civ1} presents an overview of the proposed framework. Consider a heterogeneous population of $n$ road sections. The degradation performance $y_{ij}$ for road section $i$ is observed at time points $t_{ij}$, where $i = 1, \ldots, n$ and $j = 1, \ldots, m_i$. The proposed model is,
\begin{align*}
	y_{ij} \mid h_{(i)}(t_{ij}, \thetavec_{(i)}), \sigma_{(i)}^2
	&\sim f_{(i)}( \cdot \mid h_{(i)}(t_{ij}, \thetavec_{(i)}), \sigma_{(i)}^2 ), \\
	( h_{(i)}(t, \thetavec_{(i)}),\sigma_{(i)}^2  ) \mid Q
	&\sim Q,  \quad
	Q \mid \tau, Q_0
	 \sim \text{DP}(\tau, Q_0(\cdot)),
\end{align*}
where $f_{(i)}( \cdot \mid h_{(i)}(t_{ij}, \thetavec_{(i)}), \sigma_{(i)}^2 )$ is the PDF of the sub-population that road $i$ is associated with. The sub-population degradation trend is denoted by $h_k(t, \thetavec_k)$, characterized by parameters $\thetavec_k$ and standard deviation $\sigma_k$. The parameters for all sub-populations are drawn from a distribution $Q$, which is assigned a Dirichlet process prior with parameter $\tau$ and base distribution $Q_0$. Bayesian sequential updating is performed, allowing for the integration of both historical population-level data and individual-specific observations through Bayesian inference.

\begin{figure}
\centering
\includegraphics[width=0.98\textwidth]{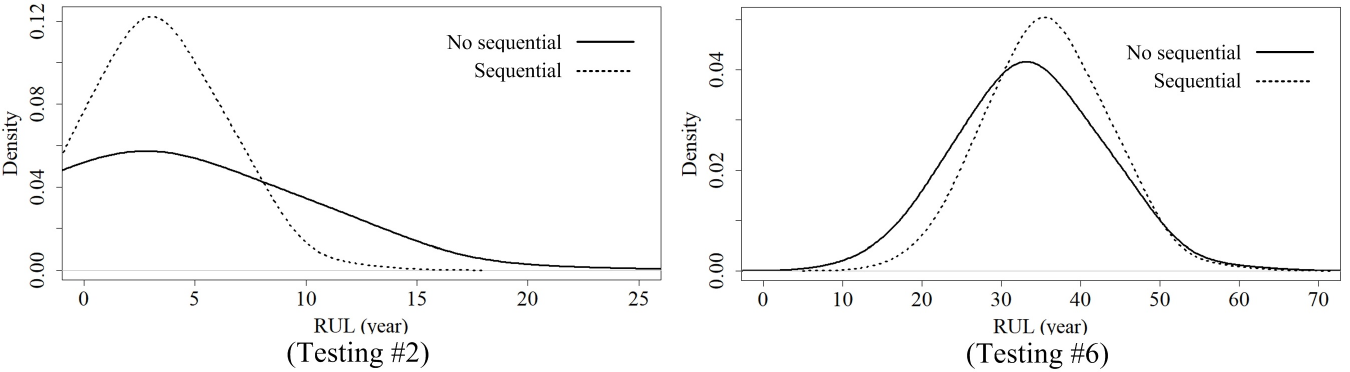}
\caption{Visualization of posterior uncertainty in predicted RUL with and without sequential updates for two representative road segments. \emph{Figure reproduced from Nguyen et al.~(2023) with permission from Wiley}.}\label{fig:civ4}
\end{figure}

To illustrate the method, \citet{nguyen2023bayesian} analyzed repeated measures degradation data from Florida road sections using annual inspection records from the Florida Department of Transportation (DOT), which track surface conditions such as cracking and patching. The international roughness index (IRI), measured in inches per mile, was used as the degradation metric. An IRI threshold of 180 inches per mile was set to define RUL, the time remaining until maintenance is needed. Figure~\ref{fig:civ4} shows how sequentially incorporating individual-level data improves RUL prediction for test units~2 and~6 by reducing posterior uncertainty and increasing precision.

\section{Broader Application Areas of Degradation Models}\label{sec:borader.application.areas}
Beyond the detailed case studies in Section~\ref{sec:applications}, we also offer broader overviews of degradation modeling in other fields. Sections~\ref{sec:solar.energy}, \ref{sec:aerospace}, and~\ref{sec:pharmaceuticals} introduce applications in the solar energy, aerospace, and pharmaceutical industries, respectively.

\subsection{Solar Energy}\label{sec:solar.energy}

In the solar energy industry, the long-term performance of photovoltaic (PV) systems is of great interest. The power output of PV modules gradually declines over time. In particular, many PV components, especially polymeric materials such as encapsulants, frontsheets, backsheets, edge sealants, and junction boxes, are susceptible to degradation and may lose their functional properties under harsh environmental conditions~\citep{Lyueta2018_solar}. Degradation models serve as valuable tools for understanding and predicting these performance declines in the solar energy sector.

Accelerated degradation testing plays a crucial role in the design stage of PV research. Given the long expected lifetimes of solar components, it is impractical to wait decades to assess their reliability. Instead, components are exposed to accelerated stress conditions such as thermal cycling, UV exposure, and humidity to induce degradation within a shorter time frame. The resulting data are then analyzed using degradation models to extrapolate long-term performance under normal operating conditions.

Degradation models are used to estimate the long-term decline in power output of PV modules in the field. These models are built using historical performance data from deployed systems, incorporating variables such as output power, irradiance, temperature, and soiling. They help predict when a panel's performance will fall below acceptable thresholds and are essential for warranty analysis. Financial institutions and utility companies often rely on these models to project long-term energy yield and evaluate the return on investment for large-scale solar farms, accounting for both gradual degradation and sudden failures such as delamination or corrosion.

\subsection{Aerospace}\label{sec:aerospace}
Degradation models serve as valuable tools for ensuring safety and reliability in the aerospace industry. During the design stage, accelerated tests are commonly conducted on aerospace components, and degradation models are subsequently applied to predict the reliability of those components. Materials and electronic systems are exposed to elevated stress conditions, such as high temperature, humidity, and vibration, to facilitate timely data collection. The data obtained under these accelerated conditions are then analyzed using degradation models to extrapolate performance and reliability under normal operating environments.

For systems deployed in the field, degradation models are primarily used for system health monitoring and condition-based maintenance~\citep{Jilkeetal2023}. Many aircraft structures, such as fuselages, wings, and landing gear, are subject to strain, fatigue, corrosion, and environmental exposure. For example, fatigue cracks may develop in turbine blades, and corrosion can occur in aluminum fuselage panels, making condition-based maintenance essential. The primary application of degradation models in this context is to predict the RUL of critical components such as jet engines, turbine blades, bearings, and fuel systems. Sensors continuously collect degradation, usage, and environmental data, which are then analyzed by degradation models to predict the RUL and predict when a performance characteristic is likely to reach a failure threshold. These predictions enable engineers to take proactive maintenance actions and prevent unexpected failures.

\subsection{Pharmaceuticals}\label{sec:pharmaceuticals}

Pharmaceutical companies utilize stability studies and degradation models throughout the drug product (DP) life cycle, excluding initial drug discovery. Within a stability study, DP quality attributes are measured over time under specific storage conditions (e.g., temperature and humidity). Statistical degradation models of the stability data are commonly used to estimate the DP degradation profile and thus characterize the DP stability over time. \cite{burdick2017statistical} and \cite{peterson2009statistics} discuss the broad applicability of stability studies, highlighting their use in DP development stages, clinical trials, new drug applications, and post-approval commitments.

Stability testing is generally conducted within the chemistry, manufacturing, and controls (CMC) area, where stability data and degradation models provide product and process understanding. Real-time stability data or model-based extrapolation beyond the observed data are also used to establish the DP shelf-life, or expiration dating period, defined as the timepoint at which DP quality attributes, such as potency or purity, reach a predefined threshold, or shelf-life specification. \cite{chow2007statistical} emphasizes the importance of shelf-life determination when stating, ``For the purpose of safety and quality assurance, most regulatory agencies such as the FDA require that an expiration dating period be indicated on the immediate container label for every drug product on the market'' (p.~1).

In 1990, representatives from US, European, and Japanese regulatory agencies and industry, formed the International Council for Harmonisation of Technical Requirements for Pharmaceuticals for Human Use (ICH) to establish global guidelines for the quality, safety, and efficacy of pharmaceutical products. The current ICH quality guidelines Q1A(R2) through Q1F focus on stability. In particular, Q1A(R2) provides guidance on the stability data package for a new DP registration application, and Q1E discuss how to use the data to propose a shelf life in the application \citep{guideline2003evaluation}. Neither guideline is heavily detailed nor prescriptive. In fact, ICH Q1A(R2) states it ``leaves sufficient flexibility'' (p. 1), and, similarly, ICH Q1E states ``the examples and references do not cover all applicable statistical approaches'' (p.~2). As such, sources like \cite{chow2007statistical} and \cite{burdick2017statistical} should be referenced for more comprehensive descriptions of experimental designs and statistical models for stability testing.

Long-term stability studies provide real-time degradation data of DP held under standard storage conditions and satisfy final regulatory requirements for shelf life. In short-term, or accelerated, stability studies, DP is exposed to elevated stress conditions, such as increased temperature and humidity, so that DP degradation can be observed significantly faster as in \cite{magari2003shelf}. Short-term stability studies are often conducted in early development phases to establish degradation rates. The relationship between degradation rates and storage conditions can then be used to extrapolate interim long-term shelf-life predictions via tools such as the Arrhenius equation \citep{chow2007statistical}.

Regardless of the type of stability study, many DP quality attributes exhibit constant, monotonically increasing or decreasing rates of change over time. In such cases, a linear regression, or zero-order kinetic model, may be fit to the data and used for shelf-life estimation. Some attributes may follow a first-order kinetic model in which initially rapid degradation becomes slower as time passes. In this case, linear regression may be applied to the log transformation of the response. Other attributes may exhibit more complex degradation profiles that require the use of nonlinear regression techniques. Beyond the linearity of the degradation profile, much debate still exists around the most appropriate statistical model for stability data, in particular the treatment of batches in cases of significant batch-to-batch variability. For example, a fixed effects model may be preferred given that most guidance only suggests a minimum of three batches. On the other hand, a mixed effects model with batch as a random effect may be favored when considering the inference space and shelf-life predictions for all future batches \citep{chow2007statistical}.

\cite{burdick2017statistical} summarizes the criticality of stability testing in the pharmaceutical industry by asserting, ``a stability study is not merely a regulatory requirement. Rather, it is a key component in acquiring scientific knowledge that supports continued quality, safety, and efficacy'' (p. 270). Statistical methods play an essential role in stability test design and analysis, yielding accurate, reliable, and valid degradation models and shelf-life predictions. Furthermore, the recent focus on quality by design and ongoing advances in pharmaceutical science, statistical methodology, and computational capability, maintain stability testing as an active research area. For instance, in 2022 ICH established an expert working group to combine and update the stability guideline series into a single guideline, ICH Q1. ICH just recently released a draft version of ICH Q1, which includes new guidance on stability modeling, as a Step 2 document for comments in April 2025.

\section{Concluding Remarks}\label{sec:concluding.remarks}

\subsection{Practical Considerations}\label{sec:practical.consideration}

As a practical guideline in degradation modeling and analysis, one should first identify the type of degradation data under consideration to select appropriate analytical approaches. For ADDT data, the GPM is typically used. For RMDT data, both GPM and SP models are available. We recommend practitioners use GPMs as they can incorporate physically motivated degradation path information. In contrast, SP models are generally not flexible enough to accommodate such physical knowledge.

It is important to keep in mind that extrapolation is often involved in degradation analysis and reliability prediction. Statistical analyses involving extrapolation are inherently challenging and should be approached with caution. A model may fit the observed data well but still perform poorly when extrapolated over time and/or accelerating variable, offering no guarantee of accuracy. Such issues are often due to overfitting. Therefore, predictions based on degradation models should be interpreted with care, and a sensitivity analysis is recommended to assess the robustness of the results.

In practice, software availability also plays a critical role, as users often rely on commercial or open-source tools to carry out analysis. Most commonly used GPMs are implemented in several software packages, which streamline the modeling process for practitioners. These challenges and available software tools will be discussed in detail in the next section.

\subsection{Software Implementations}

Here, we provide a brief overview of available software for degradation modeling and analysis. Most tools adopt the GPM framework, allowing users to select from a range of degradation path models. JMP 19.0 \citep{jmp19} offers several options within its Reliability and Survival suite, including the Repeated Measures Degradation and Destructive Degradation platforms. A general Degradation platform is also available, which supports both typical degradation analyses and stability studies. These platforms accommodate accelerated degradation data, provided only one accelerating factor is used. The Repeated Measures Degradation platform primarily employs a Bayesian approach, with users specifying priors via an interactive graphical interface. In contrast, the Destructive Degradation platform relies on maximum likelihood estimation. Both platforms support a variety of distributional assumptions and degradation path models, and they guide users through model diagnostics and lifetime prediction.

In the Weibull++ module of the ReliaSoft software \citep{reliasoft}, users can conduct degradation analysis using the GPM framework for both RMDT and ADDT data. The platform offers a user-friendly interface for fitting degradation models and generating reliability predictions based on the fitted results. It also supports accelerated degradation analysis by allowing the incorporation of accelerating variables, such as temperature, into the modeling process.

In R, several packages are available for the analysis of degradation data. The \texttt{ADDT} package \citep{ADDT} is specifically designed for analyzing ADDT data, following procedures developed for polymer degradation studies \citep{King2016}. \citet{Jinetal2017ADDT} illustrate the use of the \texttt{ADDT} package. The \texttt{chemdeg} package \citep{chemdeg_package} supports the analysis of degradation data arising from chemical kinetics. The \texttt{SPREDA} package \citep{spreda} allows for modeling degradation data with dynamic independent covariates. Despite these developments, significant efforts are still needed to create more accessible and comprehensive software tools that enable practitioners to apply the wide range of degradation models developed in the literature.

\subsection{Trends in Degradation Modeling}\label{sec:future.trends}

We highlight some emerging trends in degradation modeling research. One active area is the analysis of multivariate degradation data, where multiple degradation characteristics (DCs) are measured from the same units over time. This has become a prominent research direction in recent years (e.g., \citealt{Hajihaetal2021JQT}, and \citealt{Linetal2025}), with a recent review provided by \citet{Yietal2025RESS}. A key advantage of the multivariate approach is the ability to leverage the correlation among different DCs to improve prediction accuracy. Another trend involves modeling degradation measurements as curves rather than single or multiple discrete points, resulting in functional degradation data. \citet{Cho2024Battery} present an example of the functional approach for battery life prediction.

With the widespread deployment of sensors, sensor data have become increasingly common in modern engineering systems. For example, \citet{Jinetal2024INFORMS} discussed clustering techniques for sensor data to detect reliability-related events, while \citet{Wangetal2022index} used sensor data to construct a degradation index via variable selection methods. In recent years, the use of artificial intelligence (AI) and machine learning (ML) in degradation modeling has gained popularity, with deep learning approaches being applied to predict future degradation. However, caution is warranted when using ML for prediction, especially in settings involving extrapolation, as such models may not generalize well to new conditions. To address this concern, we advocate for physics-informed ML and deep learning approaches, which incorporate domain knowledge into the modeling process to enhance predictive performance (e.g., \citealt{Hajiha2022108677}). In parallel, Bayesian methods have also become increasingly popular for degradation analysis, particularly in the context of general path models (e.g., \citealt{meeker2021statistical}).

As AI technologies continue to advance, it becomes increasingly important to analyze the reliability of AI systems \citep{hong2023statistical}. The modeling and analysis of AI system reliability presents exciting research opportunities for statisticians \citep{Minetal2024-appliedstat}. In recent years, significant attention has been given to various aspects of AI reliability, including system-level reliability analysis \citep{Min2022AIRecurrent}, test planning in AI environments \citep{Zhengetal2025TestPlanning}, and performance evaluation of AI-generated outputs \citep{Songetal2025-coderating}. As AI systems are deployed in real-world settings, we also expect to observe performance degradation over time \citep{Zhengetal2025-datareview}. Therefore, modeling the degradation of AI performance is poised to become an important direction in the future development of degradation modeling methodologies.

\subsection{Further Readings}

A concise overview of degradation models with technical depth can be found in \citet{Meekeretal2014DegradationBookChapter}. For a more systematic and comprehensive treatment, we recommend the book by \citet{meeker2021statistical}, which covers a wide range of degradation modeling techniques, inference procedures, and practical applications. For practitioners interested in software implementations, commercial tools such as JMP and ReliaSoft Weibull++ offer built-in support for degradation analysis, along with online tutorials and instructional videos to facilitate learning. Continuous learning and adaptation are essential, given the evolving nature of the field. For those seeking to explore recent developments and emerging directions, the references provided in Section~\ref{sec:future.trends} offer a valuable starting point.

\section*{Acknowledgments}
We thank the editor and two anonymous referees for their constructive suggestions, which have significantly improved the quality of the paper. Hong's work was supported in part by the COS Dean's Discovery Fund (Award: 452021) and by the Data Science Faculty Fellowship (Award: 452118), both at Virginia Tech. GPT-4o was used as a grammar checking tool in the preparation of this manuscript.

\def\baselinestretch{1.0}
\begin{small}

\end{small}


\begin{thebibliography}{}

\bibitem[\protect\citeauthoryear{{American Society of Civil
  Engineers}}{{American Society of Civil Engineers}}{2025}]{mingy1}
{American Society of Civil Engineers} (2025).
\newblock 2025 report card for {America}'s infrastructure.
\newblock \url{https://infrastructurereportcard.org}.
\newblock Accessed: 2025-04-14.

\bibitem[\protect\citeauthoryear{Bi}{Bi}{2024}]{bi2024degradation}
Bi, S. (2024).
\newblock Degradation performance analysis and reliability quantitative
  evaluation technology of electronic equipment.
\newblock In {\em 2024 International Conference on Computing, Robotics and
  System Sciences (ICRSS)}, pp.\  129--133. IEEE.

\bibitem[\protect\citeauthoryear{Burdick, LeBlond, Pfahler, Quiroz, Sidor,
  Vukovinsky, and Zhang}{Burdick et~al.}{2017}]{burdick2017statistical}
Burdick, R.~K., D.~J. LeBlond, L.~B. Pfahler, J.~Quiroz, L.~Sidor,
  K.~Vukovinsky, and L.~Zhang (2017).
\newblock {\em Statistical applications for chemistry, manufacturing and
  controls (CMC) in the pharmaceutical industry}, Volume~10.
\newblock Springer.

\bibitem[\protect\citeauthoryear{Cho, Do, Du, and Hong}{Cho
  et~al.}{2024}]{Cho2024Battery}
Cho, Y., Q.~Do, P.~Du, and Y.~Hong (2024).
\newblock Reliability study of battery lives: A functional degradation analysis
  approach.
\newblock {\em The Annals of Applied Statistics\/}~{\em 18\/}(4), 3185--3204.

\bibitem[\protect\citeauthoryear{Chow}{Chow}{2007}]{chow2007statistical}
Chow, S.-C. (2007).
\newblock {\em Statistical design and analysis of stability studies}.
\newblock Chapman and Hall/CRC.

\bibitem[\protect\citeauthoryear{Condra}{Condra}{1993}]{Condra1993}
Condra, L.~W. (1993).
\newblock {\em Reliability Improvement with Design of Experiments}.
\newblock New York: Marcel Deller.

\bibitem[\protect\citeauthoryear{Dowling}{Dowling}{2013}]{Dowling1993}
Dowling, N.~E. (2013).
\newblock {\em Mechanical behavior of materials: Engineering methods for
  deformation, fracture, and fatigue\/} (4th ed.).
\newblock London, UK: Pearson.

\bibitem[\protect\citeauthoryear{Duan, Hong, Meeker, Stanley, and Gu}{Duan
  et~al.}{2017}]{Duanetal2017}
Duan, Y., Y.~Hong, W.~Meeker, D.~Stanley, and X.~Gu (2017).
\newblock Photodegradation modeling based on laboratory accelerated test data
  and predictions under outdoor weathering for polymeric materials.
\newblock {\em The Annals of Applied Statistics\/}~{\em 11}, 2052--2079.

\bibitem[\protect\citeauthoryear{Escobar, Meeker, Kugler, and Kramer}{Escobar
  et~al.}{2003}]{EscobarMeekerKuglerKramer2003}
Escobar, L.~A., W.~Q. Meeker, D.~L. Kugler, and L.~L. Kramer (2003).
\newblock Accelerated destructive degradation tests: Data, models, and
  analysis.
\newblock In B.~H. Lindqvist and K.~A. Doksum (Eds.), {\em Mathematical and
  Statistical Methods in Reliability}. Singapore: World Scientific Publishing
  Company.

\bibitem[\protect\citeauthoryear{Esslinger, Kieselbach, Koller, and
  Weisse}{Esslinger et~al.}{2004}]{esslinger2004railway}
Esslinger, V., R.~Kieselbach, R.~Koller, and B.~Weisse (2004).
\newblock The railway accident of eschede--technical background.
\newblock {\em Engineering Failure Analysis\/}~{\em 11\/}(4), 515--535.

\bibitem[\protect\citeauthoryear{Focke, {van der Westhuizen}, Musee, and
  Loots}{Focke et~al.}{2017}]{Fockeetal2017}
Focke, W.~W., I.~{van der Westhuizen}, N.~Musee, and M.~T. Loots (2017).
\newblock Kinetic interpretation of log-logistic dose-time response curves.
\newblock {\em Scientific Reports\/}~{\em 7\/}(1), 2234.

\bibitem[\protect\citeauthoryear{Hajiha, Liu, and Hong}{Hajiha
  et~al.}{2021}]{Hajihaetal2021JQT}
Hajiha, M., X.~Liu, and Y.~Hong (2021).
\newblock Degradation under dynamic operating conditions: Modeling, competing
  processes and applications.
\newblock {\em Journal of Quality Technology\/}~{\em 53\/}(4), 347--368.

\bibitem[\protect\citeauthoryear{Hajiha, Liu, Lee, and Ramin}{Hajiha
  et~al.}{2022}]{Hajiha2022108677}
Hajiha, M., X.~Liu, Y.~M. Lee, and M.~Ramin (2022).
\newblock A physics-regularized data-driven approach for health prognostics of
  complex engineered systems with dependent health states.
\newblock {\em Reliability Engineering \ System Safety\/}~{\em 226}, 108677.

\bibitem[\protect\citeauthoryear{Hoffman and Gelman}{Hoffman and
  Gelman}{2014}]{hoffman2014no}
Hoffman, M.~D. and A.~Gelman (2014).
\newblock The {No-U-Turn} sampler: adaptively setting path lengths in
  {Hamiltonian} {Monte} {Carlo}.
\newblock {\em Journal of Machine Learning Research\/}~{\em 15}, 1593--1623.

\bibitem[\protect\citeauthoryear{Hong, Duan, Meeker, Stanley, and Gu}{Hong
  et~al.}{2015}]{HongDuanetal2015}
Hong, Y., Y.~Duan, W.~Q. Meeker, D.~L. Stanley, and X.~Gu (2015).
\newblock Statistical methods for degradation data with dynamic covariates
  information and an application to outdoor weathering data.
\newblock {\em Technometrics\/}~{\em 57}, 180--193.

\bibitem[\protect\citeauthoryear{Hong, Lian, Xu, Min, Wang, Freeman, and
  Deng}{Hong et~al.}{2023}]{hong2023statistical}
Hong, Y., J.~Lian, L.~Xu, J.~Min, Y.~Wang, L.~J. Freeman, and X.~Deng (2023).
\newblock Statistical perspectives on reliability of artificial intelligence
  systems.
\newblock {\em Quality Engineering\/}~{\em 35\/}(1), 56--78.

\bibitem[\protect\citeauthoryear{Hong, Xie, Jin, and King}{Hong
  et~al.}{2016}]{ADDT}
Hong, Y., Y.~Xie, Z.~Jin, and C.~King (2016).
\newblock {\em {ADDT}: Analysis of Accelerated Destructive Degradation Test
  Data}.
\newblock R package version 2.0; https://CRAN.R-project.org/package=ADDT.

\bibitem[\protect\citeauthoryear{Hong, Xie, and Xu}{Hong et~al.}{2018}]{spreda}
Hong, Y., Y.~Xie, and Z.~Xu (2018).
\newblock {\em {SPREDA}: Statistical Package for Reliability Data Analysis}.
\newblock R package version 1.1.

\bibitem[\protect\citeauthoryear{ICH}{ICH}{2003}]{guideline2003evaluation}
ICH (2003).
\newblock Evaluation for stability data.
\newblock {\em ICH Q1E. ICH\/}~{\em 6}, 1--19.

\bibitem[\protect\citeauthoryear{Jilke, Raddatz, and Wende}{Jilke
  et~al.}{2023}]{Jilkeetal2023}
Jilke, L., F.~Raddatz, and G.~Wende (2023).
\newblock Investigation of degradation modeling for aircraft structures: A
  systematic literature review.
\newblock {\em AIAA AVIATION 2023 Forum, DOI: 10.2514/6.2023-3550\/}.

\bibitem[\protect\citeauthoryear{Jin, Min, Hong, Du, and Yang}{Jin
  et~al.}{2024}]{Jinetal2024INFORMS}
Jin, Z., J.~Min, Y.~Hong, P.~Du, and Q.~Yang (2024).
\newblock Multivariate functional clustering with variable selection and
  application to sensor data from engineering systems.
\newblock {\em {INFORMS} Journal on Data Science\/}~{\em 3\/}(2), 203--218.

\bibitem[\protect\citeauthoryear{Jin, Xie, Hong, and Van~Mullekom}{Jin
  et~al.}{2017}]{Jinetal2017ADDT}
Jin, Z., Y.~Xie, Y.~Hong, and J.~H. Van~Mullekom (2017).
\newblock {\em {ADDT}: An {R} Package for Analysis of Accelerated Destructive
  Degradation Test Data}.
\newblock Singapore: Springer Singapore.

\bibitem[\protect\citeauthoryear{JMP}{JMP}{2025}]{jmp19}
JMP (2025).
\newblock {JMP}\textsuperscript{\textregistered} statistical discovery
  software, version 19.0.

\bibitem[\protect\citeauthoryear{Kaplan and Meier}{Kaplan and
  Meier}{1958}]{KaplanMeier1958}
Kaplan, E.~L. and P.~Meier (1958).
\newblock Nonparametric estimation from incomplete observations.
\newblock {\em Journal of the American Statistical Association\/}~{\em 53},
  457--481.

\bibitem[\protect\citeauthoryear{King, Xie, Hong, Mullekom, DeHart, and
  DeFeo}{King et~al.}{2018}]{King2016}
King, C.~B., Y.~Xie, Y.~Hong, J.~H.~V. Mullekom, S.~P. DeHart, and P.~A. DeFeo
  (2018).
\newblock A comparison of traditional and maximum likelihood approaches to
  estimating thermal indices for polymeric materials.
\newblock {\em Journal of Quality Technology\/}~{\em 50}, 117--129.

\bibitem[\protect\citeauthoryear{Kozin and Bogdanoff}{Kozin and
  Bogdanoff}{1981}]{kozin1981critical}
Kozin, F. and J.~Bogdanoff (1981).
\newblock A critical analysis of some probabilistic models of fatigue crack
  growth.
\newblock {\em Engineering Fracture Mechanics\/}~{\em 14\/}(1), 59--89.

\bibitem[\protect\citeauthoryear{Lawless and Crowder}{Lawless and
  Crowder}{2004}]{LawlessCrowder2004}
Lawless, J. and M.~Crowder (2004).
\newblock Covariates and random effects in a gamma process model with
  application to degradation and failure.
\newblock {\em Lifetime Data Analysis\/}~{\em 10}, 213--227.

\bibitem[\protect\citeauthoryear{Lin, Liu, Xiang, and Hong}{Lin
  et~al.}{2025}]{Linetal2025}
Lin, Z., X.~Liu, Y.~Xiang, and Y.~Hong (2025).
\newblock Modeling multivariate degradation data with dynamic covariates under
  a {Bayesian} framework.
\newblock {\em Reliability Engineering \& System Safety\/}~{\em 261}, 111115.

\bibitem[\protect\citeauthoryear{Lu and Meeker}{Lu and
  Meeker}{1993}]{LuMeeker1993}
Lu, C.~J. and W.~Q. Meeker (1993).
\newblock Using degradation measures to estimate a time-to-failure
  distribution.
\newblock {\em Technometrics\/}~{\em 34}, 161--174.

\bibitem[\protect\citeauthoryear{Lu, Meeker, and Escobar}{Lu
  et~al.}{1996}]{lu1996comparison}
Lu, C.~J., W.~Q. Meeker, and L.~A. Escobar (1996).
\newblock A comparison of degradation and failure-time analysis methods for
  estimating a time-to-failure distribution.
\newblock {\em Statistica Sinica\/}~{\em 6\/}(3), 531--546.

\bibitem[\protect\citeauthoryear{Lu, Wang, Hong, and Ye}{Lu
  et~al.}{2021}]{lu2021general}
Lu, L., B.~Wang, Y.~Hong, and Z.~Ye (2021).
\newblock General path models for degradation data with multiple
  characteristics and covariates.
\newblock {\em Technometrics\/}~{\em 63\/}(3), 354--369.

\bibitem[\protect\citeauthoryear{Lyu, Kim, and Gu}{Lyu
  et~al.}{2018}]{Lyueta2018_solar}
Lyu, Y., J.~H. Kim, and X.~Gu (2018).
\newblock Developing methodology for service life prediction of {PV} materials:
  Quantitative effects of light intensity and wavelength on discoloration of a
  glass/{EVA}/{PPE} laminate.
\newblock {\em Solar Energy\/}~{\em 174}, 515--526.

\bibitem[\protect\citeauthoryear{Magari}{Magari}{2003}]{magari2003shelf}
Magari, R.~T. (2003).
\newblock Assessing shelf life using real-time and accelerated stability tests.
\newblock {\em BioPharm International\/}~{\em 16\/}(11), 36--48.

\bibitem[\protect\citeauthoryear{Meeker, Hong, and Escobar}{Meeker
  et~al.}{2011}]{Meekeretal2014DegradationBookChapter}
Meeker, W., Y.~Hong, and L.~Escobar (2011).
\newblock Degradation models and analyses.
\newblock In {\em Encyclopedia of Statistical Sciences}, pp.\  1--23. John
  Wiley \& Sons, Ltd.

\bibitem[\protect\citeauthoryear{Meeker and Escobar}{Meeker and
  Escobar}{1998}]{meeker1998statistical}
Meeker, W.~Q. and L.~A. Escobar (1998).
\newblock {\em Statistical methods for reliability data}.
\newblock John Wiley \& Sons.

\bibitem[\protect\citeauthoryear{Meeker, Escobar, and Lu}{Meeker
  et~al.}{1998}]{MeekerEscobarLu1998}
Meeker, W.~Q., L.~A. Escobar, and C.~J. Lu (1998).
\newblock Accelerated degradation tests: modeling and analysis.
\newblock {\em Technometrics\/}~{\em 40}, 89--99.

\bibitem[\protect\citeauthoryear{Meeker, Escobar, and Pascual}{Meeker
  et~al.}{2022}]{meeker2021statistical}
Meeker, W.~Q., L.~A. Escobar, and F.~G. Pascual (2022).
\newblock {\em Statistical methods for reliability data\/} (2nd ed.).
\newblock John Wiley \& Sons.

\bibitem[\protect\citeauthoryear{Migliorini, Fierri, Zoccatelli, and
  Chignola}{Migliorini et~al.}{2024}]{chemdeg_package}
Migliorini, M., I.~Fierri, G.~Zoccatelli, and R.~Chignola (2024).
\newblock Chemdeg, an {R} package for the analysis of foods isothermal
  degradation kinetics.
\newblock {\em Journal of Food Engineering\/}~{\em 363\/}(0260-8774), 111778.

\bibitem[\protect\citeauthoryear{Min, Hong, King, and Meeker}{Min
  et~al.}{2022}]{Min2022AIRecurrent}
Min, J., Y.~Hong, C.~B. King, and W.~Q. Meeker (2022).
\newblock Reliability analysis of artificial intelligence systems using
  recurrent events data from autonomous vehicles.
\newblock {\em Journal of the Royal Statistical Society: Series C (Applied
  Statistics)\/}~{\em 71\/}(4), 987--1013.

\bibitem[\protect\citeauthoryear{Min, Song, Zheng, King, Deng, and Hong}{Min
  et~al.}{2024}]{Minetal2024-appliedstat}
Min, J., X.~Song, S.~Zheng, C.~B. King, X.~Deng, and Y.~Hong (2024).
\newblock Applied statistics in the era of artificial intelligence: A review
  and vision.
\newblock {\em arXiv: 2303.16369\/}.

\bibitem[\protect\citeauthoryear{Nair}{Nair}{1984}]{Nair1984}
Nair, V.~N. (1984).
\newblock Confidence bands for survival functions with censored data: a
  comparative study.
\newblock {\em Technometrics\/}~{\em 26}, 265--275.

\bibitem[\protect\citeauthoryear{Nguyen, Sun, Lu, Zhang, and Li}{Nguyen
  et~al.}{2023}]{nguyen2023bayesian}
Nguyen, H., X.~Sun, Q.~Lu, Q.~Zhang, and M.~Li (2023).
\newblock Bayesian heterogeneous degradation performance modeling with an
  unknown number of sub-populations.
\newblock {\em Quality and Reliability Engineering International\/}~{\em
  39\/}(7), 2686--2705.

\bibitem[\protect\citeauthoryear{Peng}{Peng}{2016}]{Peng2016}
Peng, C.-Y. (2016).
\newblock Inverse {Gaussian} processes with random effects and explanatory
  variables for degradation data.
\newblock {\em Technometrics\/}~{\em 57}, 100--111.

\bibitem[\protect\citeauthoryear{Peng, Li, Yang, Mi, and Huang}{Peng
  et~al.}{2017}]{7803533}
Peng, W., Y.-F. Li, Y.-J. Yang, J.~Mi, and H.-Z. Huang (2017).
\newblock Bayesian degradation analysis with inverse gaussian process models
  under time-varying degradation rates.
\newblock {\em IEEE Transactions on Reliability\/}~{\em 66\/}(1), 84--96.

\bibitem[\protect\citeauthoryear{Peterson, Snee, McAllister, Schofield, and
  Carella}{Peterson et~al.}{2009}]{peterson2009statistics}
Peterson, J.~J., R.~D. Snee, P.~R. McAllister, T.~L. Schofield, and A.~J.
  Carella (2009).
\newblock Statistics in pharmaceutical development and manufacturing.
\newblock {\em Journal of Quality Technology\/}~{\em 41\/}(2), 111--134.

\bibitem[\protect\citeauthoryear{Pinheiro and Bates}{Pinheiro and
  Bates}{2006}]{pinheiro2006nlme}
Pinheiro, J.~C. and D.~M. Bates (2006).
\newblock {\em Mixed-Effects Models in {S} and {S-PLUS}}.
\newblock New York: Springer.
\newblock R package version 3.1-164.

\bibitem[\protect\citeauthoryear{{ReliaSoft Corporation}}{{ReliaSoft
  Corporation}}{2024}]{reliasoft}
{ReliaSoft Corporation} (2024).
\newblock Reliasoft software suite.
\newblock \texttt{https://www.reliasoft.com/}.

\bibitem[\protect\citeauthoryear{Song, Xie, Lee, Chen, Clark, He, He, Min,
  Zhang, Zheng, Zhang, Deng, and Hong}{Song
  et~al.}{2025}]{Songetal2025-coderating}
Song, X., K.~Xie, L.~Lee, R.~Chen, J.~M. Clark, H.~He, H.~He, J.~Min, X.~Zhang,
  S.~Zheng, Z.~Zhang, X.~Deng, and Y.~Hong (2025).
\newblock Performance evaluation of large language models in statistical
  programming.
\newblock {\em arXiv preprint arXiv:2502.13117\/}.

\bibitem[\protect\citeauthoryear{UL746B}{UL746B}{2001}]{UL746B}
UL746B (2001).
\newblock {\em Polymeric Materials - Long Term Property Evaluations, UL 746B}.
\newblock Underwriters Laboratories, Incorporated.

\bibitem[\protect\citeauthoryear{Wang and Xu}{Wang and Xu}{2010}]{WangXu2010}
Wang, X. and D.~Xu (2010).
\newblock An inverse \protect{Gaussian} process model for degradation data.
\newblock {\em Technometrics\/}~{\em 52}, 188--197.

\bibitem[\protect\citeauthoryear{Wang, Lee, Hong, and Deng}{Wang
  et~al.}{2022}]{Wangetal2022index}
Wang, Y., I.-C. Lee, Y.~Hong, and X.~Deng (2022).
\newblock Building degradation index with variable selection for multivariate
  sensory data.
\newblock {\em Reliability Engineering \& System Safety\/}~{\em 227}, 108704.

\bibitem[\protect\citeauthoryear{Whitmore}{Whitmore}{1995}]{Whitmore1995}
Whitmore, G.~A. (1995).
\newblock Estimation degradation by a \protect{Wiener} diffusion process
  subject to measurement error.
\newblock {\em Lifetime Data Analysis\/}~{\em 1}, 307--319.

\bibitem[\protect\citeauthoryear{Withey}{Withey}{1997}]{withey1997fatigue}
Withey, P. (1997).
\newblock Fatigue failure of the de {Havilland} comet {I}.
\newblock {\em Engineering Failure Analysis\/}~{\em 4\/}(2), 147--154.

\bibitem[\protect\citeauthoryear{Xie, Jin, Hong, and Van~Mullekom}{Xie
  et~al.}{2017}]{Xieetal2017ThermalIndex}
Xie, Y., Z.~Jin, Y.~Hong, and J.~H. Van~Mullekom (2017).
\newblock Statistical methods for thermal index estimation based on accelerated
  destructive degradation test data.
\newblock In {\em Statistical Modeling for Degradation Data}, pp.\  231--251.
  Springer Singapore.

\bibitem[\protect\citeauthoryear{Xie, King, Hong, and Yang}{Xie
  et~al.}{2018}]{Xieetal2018}
Xie, Y., C.~B. King, Y.~Hong, and Q.~Yang (2018).
\newblock Semi-parametric models for accelerated destructive degradation test
  data analysis.
\newblock {\em Technometrics\/}~{\em 60}, 222--234.

\bibitem[\protect\citeauthoryear{Ye and Xie}{Ye and
  Xie}{2015}]{YeXie2015DegradationReview}
Ye, Z.-S. and M.~Xie (2015).
\newblock Stochastic modelling and analysis of degradation for highly reliable
  products.
\newblock {\em Applied Stochastic Models in Business and Industry\/}~{\em
  31\/}(1), 16--32.

\bibitem[\protect\citeauthoryear{Yi, Zhang, Wang, Zhang, and Zhai}{Yi
  et~al.}{2025}]{Yietal2025RESS}
Yi, H., W.~Zhang, G.~Wang, X.~Zhang, and Q.~Zhai (2025).
\newblock Statistical multivariate degradation modeling -- a systematic review.
\newblock {\em Reliability Engineering \& System Safety\/}, 111286.

\bibitem[\protect\citeauthoryear{Zhai, Li, and Chen}{Zhai
  et~al.}{2024}]{zhai2024modeling}
Zhai, Q., Y.~Li, and P.~Chen (2024).
\newblock Modeling product degradation with heterogeneity: A general
  random-effects wiener process approach.
\newblock {\em IISE Transactions\/}~{\em 57\/}(12), 1--14.

\bibitem[\protect\citeauthoryear{Zhai and Ye}{Zhai and Ye}{2023}]{Zhai03072023}
Zhai, Q. and Z.-S. Ye (2023).
\newblock A multivariate stochastic degradation model for dependent performance
  characteristics.
\newblock {\em Technometrics\/}~{\em 65\/}(3), 315--327.

\bibitem[\protect\citeauthoryear{Zheng, Clark, Salboukh, Silva, da~Mata, Pan,
  Min, Lian, King, Fiondella, Liu, Deng, and Hong}{Zheng
  et~al.}{}]{Zhengetal2025-datareview}
Zheng, S., J.~M. Clark, F.~Salboukh, P.~Silva, K.~da~Mata, F.~Pan, J.~Min,
  J.~Lian, C.~B. King, L.~Fiondella, J.~Liu, X.~Deng, and Y.~Hong.
\newblock {DR-AIR}: A data repository bridging the research gap in ai
  reliability.
\newblock {\em Quality Engineering, DOI: 10.1080/08982112.2025.2539834\/}.

\bibitem[\protect\citeauthoryear{Zheng, Lu, Hong, and Liu}{Zheng
  et~al.}{2026}]{Zhengetal2025TestPlanning}
Zheng, S., L.~Lu, Y.~Hong, and J.~Liu (2026).
\newblock Planning reliability assurance tests for autonomous vehicles based on
  disengagement events data.
\newblock {\em IISE Transactions\/}~{\em 58\/}(2), 131--146.

\end{thebibliography}
\end{document}